\documentclass[aps,pra,twocolumn]{revtex4-2}
\usepackage{amssymb}
\usepackage{amsmath}
\usepackage{feynmf}
\usepackage{makeidx}
\usepackage{graphicx,color}
\usepackage{hyperref}
\usepackage{subfigure}
\usepackage{booktabs}
\usepackage{array}
\usepackage{multirow}
\usepackage{float}

\begin{document}

\title{Tuning the Quasi-bound States of Double-barrier Structures: Insights from Resonant
	Tunneling Spectra}
\date{\today}
\author{ Wei Li}
\altaffiliation{Corresponding author: wliustc@aust.edu.cn}
\affiliation{Center for Fundamental Physics, School of Mechanics and opticelectrical Physics, Anhui University of Science and Technology, Huainan, Anhui 232001, People's Republic of China}

\author{Yong Yang}

\altaffiliation{Corresponding author: yyanglab@issp.ac.cn}
\affiliation{Key Lab of Photovoltaic and Energy Conservation Materials, Institute of Solid State Physics, HFIPS, Chinese
	Academy of Sciences, Hefei 230031, China}
\affiliation{Science Island Branch of Graduate School, University of Science and Technology of China, Hefei 230026,
	China}

\begin{abstract}
	In this work, we study the resonant tunneling (RT) of electrons and H atoms in double-barrier (DB) systems. Our numerical calculations directly verify the correspondence between the resonant tunneling energies and the energy levels of quasi-bound states (QBS) within the double barriers. Based on this, in-depth analyses are carried out on the modulation of QBS energy levels and numbers which show step variation with the inter-barrier spacing. The mathematical criterion for the existence of QBS is derived, and the impacts of the barrier width and barrier height on QBS levels are investigated. Taking the rectangular double-barrier as an example, we have studied the manipulation of electronic structures and optical properties of the inter-barrier region (quasi-potential well) by tuning the inter-barrier spacing (width of quasi-potential well). Atom-like optical absorption features are found in the range of infrared to visible spectrum, which can be continuously tuned by the variation of quasi-potential well width. The potential application of double-barrier nanostructures in ultrahigh-precision detection of electromagnetic radiations is demonstrated. 
\end{abstract}


\maketitle


\section{Introduction}

Resonant tunneling (RT) is a unique phenomenon in quantum tunneling, specifically occurring in a double-barrier (DB) system. In this scenario, an incident particle can traverse the barriers with a probability of 100\%. The exploration of RT began with the foundational theoretical and experimental work by Tsu, Esaki, Chang, and others in the 1970s \cite{tsu1973Tunneling, chang1974Resonant, esaki1974New}, gaining considerable attention in the 1980s \cite{luryi1985Frequency, tsuchiya1985Room, stone1985Effect, capasso1986Resonant, hauge1987Transmission, weil1987Equivalence, jonson1987Effect, zaslavsky1988Resonant}. Research on RT continues today \cite{encomendero2017New, encomendero2019Broken, tao2019Coherent, zangwill2022Dynamic, bi2021Quantum, bi2021Atomic, yang2024Penetration, lin2024Resonant}, largely due to its applications in microelectronic devices such as resonant tunneling diodes \cite{mandra2014Helium, encomendero2017New, growden2018431, saito2019Tunneling}. Recent studies have investigated dynamic RT through quasi-bound superstates (QBSS) generated by oscillating delta-function potentials \cite{zangwill2022Dynamic}. 

The prevalent conceptual framework for RT suggests that the energy of the incident particle aligns with one of the quasi-bound states (QBS) energy levels formed within the DBs, leading to resonance and constructive interference of the wave functions, thus maximizing tunneling probability \cite{chang1974Resonant, esaki1974New, ricco1984Physics, capasso1986Resonant, zangwill2020Spatial, zangwill2022Dynamic}. However, this picture remains largely hypothetical and has yet to be substantiated through numerical or experimental verification. In a related work, one of the authors has performed a systematic investigation of quantum tunneling through DBs of arbitrary shape \cite{yang2024Penetration}, establishing general conditions for RT. This work demonstrates that RT can be realized for any particle with incident energy less than the barrier height by adjusting the distance between the barriers. This implies that continuous tuning of the QBS energy levels is possible through manipulation of the barrier distance, effectively modifying the width of the quasi-potential well.

Based on a number of DB systems, we have meticulously examined in this paper the RT of electrons, as well as the RT of H atoms whose tunneling effects have been demonstrated experimentally \cite{lauhon2000Direct, meng2015Direct, guo2016Nuclear}. Comprehensive numerical simulations have validated the one-to-one correspondence between the RT energies and the QBS energy levels within the double-barrier region. Utilizing this insight, we conducted a detailed analysis of how varying the distance between the barriers influences the position and abundance of QBS levels. Additionally, we elucidated the mathematical conditions necessary for the existence of these energy levels and explored the impacts of barrier width and height variations.

Focusing on rectangular DB systems, we investigated the effects of inter-barrier spacing (the width of the quasi-potential well) on the electronic structures and optical properties of the region. This examination highlights the free-atom-like electronic and optical features of QBS in DB nanostructures, which enable potential applications in ultrahigh-precision detection of electromagnetic radiation, underscoring their transformative potential in the fields of nanoelectronics and nanophotonics.

The rest of this paper is organized as follows. Section \ref{sec:mechanism} demonstrates numerically the correspondence between the RT energies and QBS levels. Section \ref{sec:wdith} elucidates how the spacing of potential barriers can modulate the energy levels of QBS and deduces the conditions that must be met with for RT to take place. Section \ref{sec:opab} examines the optical properties of the quasi-potential well region, with a particular focus on the effects of varying the inter-barrier spacing (width of quasi-potential well) and its potential applications in the ultrahigh-precision detection of electromagnetic waves within the infrared spectrum.

\section{correspondence of the Tunneling Spectra and the Quasi-bound States}\label{sec:mechanism}

\begin{figure*}[htp]
	\centering		
	\includegraphics[width=0.95\linewidth]{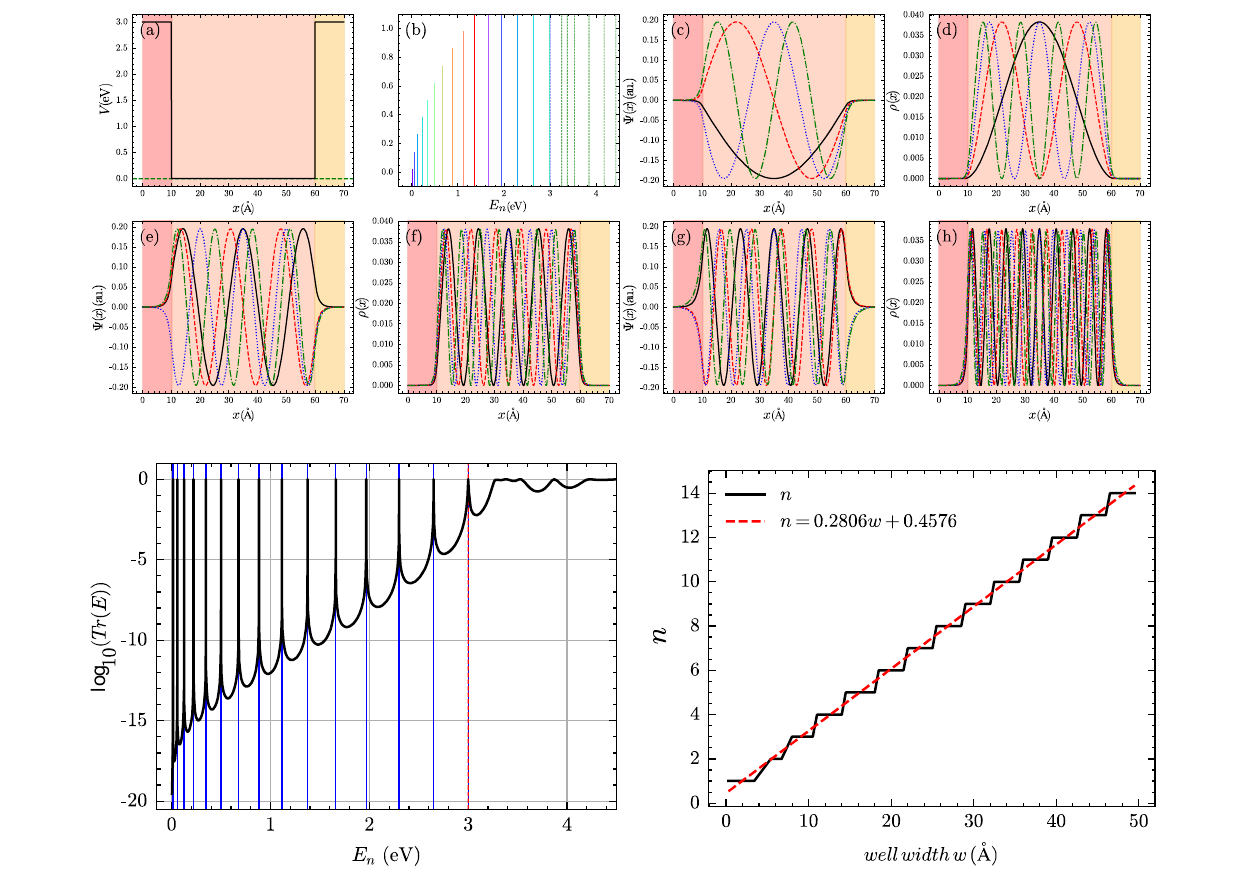}
	
	\caption{Correspondence of the quasi-bound states (QBS) energy levels of electrons (15 levels) and the resonant tunneling (RT) energies for electrons. The parameters for the
		DBs system are: barrier height $V_0$ = 3 $\rm{eV}$, barrier width $a = 10\, $\AA, well width $w = 50\,$ \AA. Top panels (a-h): The results derived from the exact diagonalization method.
		(a) A schematic diagram of the double-barrier system, with the quasi-potential well shaded by shallow pink; (b) Distribution of energy levels, with a blue dashed lines indicating the positions of the barrier heights $V_0 = 3.0\,\rm{eV}$. For graphical clarity, the data lines representing distinct QBS levels  along the  the vertical axis, have been scaled by their respective energy level coefficients.  (c, e, g) Distribution of wave functions for  $n=1-12$ and the corresponding probability distributions in (d, f, h), respectively. Bottom panels: On the left, the transmission spectra calculated by the transfer matrix method (TMM), with the transmission probability plotted on a logarithmic scale.
		The energy levels are list in Table \ref{tab:ee_eis_trs}. The bottom right panel, shows the relationship between the number of QBS and the barrier spacing (the width of the well) $w$. The red straight dashed line represents the linear fit $n = A w + B$ with $A = 0.2806,\, B= 0.4576$. }
	\label{fig:ee_eis_trs}
\end{figure*}
In this section, we study the one-to-one correspondence between the QBS energy levels and the RT energy levels, using rectangular double-barriers (DBs) as the model systems. Numerically, the QBS energy levels were obtained by solving the Schrödinger equation in one-dimensional (1D) systems 	$[ -\frac{\hbar^2}{2m}\frac{\partial^2}{\partial x^2}+V(x)]\psi(x) = E\psi(x)$
where $m$ denotes the particle mass, such as the electron or a hydrogen atom considered in this study, $\hbar$ is the reduced Planck constant, and $V(x)$ represents the potential function. We consider a symmetrical DB quantum-well structure illustrated in Fig. \ref{fig:ee_eis_trs}(a), which can be easily generalized to asymmetrical configurations and will not be considered here for simplicity. A potential well of width $w$ is located between two barriers each of which with a barrier width $a$, and barrier height $V_0$, respectively. The wave-functions $[\psi(x)]$ and eigenvalues ($E$) related to QBS can be obtained by exact diagonalization of the Schrödinger equation in real space, subjected to the boundary condition of $\psi[\pm L] = 0$, with $x = \pm L$ being the edge sites of the DB.

The quantum tunneling across double-barriers of any shape can be quantified using the transfer matrix method (TMM), a powerful technique for studying the transmission properties in nonperiodic systems \cite{tsu1973Tunneling, bi2021Atomic, bi2021Quantum, yang2024Penetration, mello1988Macroscopic, pereyra1998Resonant, pereyra2002Theory, pereyra2022Transfer}. For the propagation of a quantum particle across a single barrier $V(x)$ with compact support (the intrinsic property of physical interactions), the transmitted and reflected amplitudes ($A_L,\, B_L;\, A_R, B_R$) of the wave functions ($\psi_L,\, \psi_R$) may be related by a transfer matrix (denoted by $M$) as follows \cite{bi2021Atomic, bi2021Quantum, yang2024Penetration}
\begin{eqnarray}\label{eq:transfer_matrix}
	\left( \begin{array}{c}
		A_R \\
		B_R
	\end{array}\right) = M\left( \begin{array}{c}
		A_L \\
		B_L
	\end{array}\right) = \left( \begin{array}{cc}
		m_{11} & m_{12} \\
		m_{21} & m_{22}
	\end{array}\right) \left( \begin{array}{c}
		A_L \\
		B_L
	\end{array}\right).
\end{eqnarray}
The incoming wave function (with incident energy $E$) is expressed by $\psi_L = A_L e^{ikx} + B_L e^{-ikx}$, and the outgoing wave function is $\psi_R = A_R e^{ikx} + B_R e^{-ikx}$, where $k = \sqrt{2m E /\hbar ^2}$, and $m$ is the particle mass. The determinant $|M| = 1$, for systems where time-reversal symmetry preserves, and the transmission coefficient is given by $T = \frac{1}{|m_{11}|^2}$. In general, the matrix elements $m_{i j}$ (the subscripts i, j = 1, 2) are complex numbers and obey the conjugate relations $m_{11} = m_{22}^{*} $ and $m_{12} = m_{21}^{*}$. To numerically calculate the transmission coefficient of a quantum particle, the entire barrier is sliced to obtain a chain of rectangular potential barriers ($V_1$, $V_2$, …, $V_j$, …, $V_{n-1}$, $V_{n}$). Transmission through each of these rectangular potential barriers is similarly described using  the aforementioned Eq. (\ref{eq:transfer_matrix}), via a transfer matrix ($M_{j}$). The global transfer matrix $M$ is obtained as follows:
\begin{eqnarray}\label{eq:transfer_matrix_3}
	M = \prod_{j=n... 1}M_{j}= \left( \begin{array}{cc}
		m_{11} & m_{12} \\
		m_{21} & m_{22}
	\end{array}\right)
\end{eqnarray}
The transmission coefficient is calculated by
\begin{eqnarray}
	T_{r}(E) &=& \left|\frac{A_R}{A_L}\right|^2
	\times \frac{K_R}{K_L}
	=\frac{|M|^2}{|m_{22}|^2}
	\times \frac{K_R}{K_L}
\end{eqnarray}
where $A_L,\, A_R;\, K_L,\, K_R$ are the incident amplitude, the transmitted amplitude, the incident wave vector and the transmitted wave vector, respectively; $|M|$ is the determinant of $M$. It follows that the condition when the transmission coefficient $ T_{r}(E) = 1$ corresponds to RT.

The studies on the one-to-one correspondence of RT and QBS levels in DB systems have been carried out for electrons and hydrogen (H) atoms, to demonstrate the universality of the picture and to highlight the effects of particle mass. The electron QBS levels and wave functions, as determined by the exact diagonalization method \cite{landau2007Computational}, and the RT levels calculated by the TMM are presented in Fig. \ref{fig:ee_eis_trs}. It is clearly seen that the QBS are predominantly confined within the quasi-potential well, with only a small portion extending into the barrier region [Figs. \ref{fig:ee_eis_trs}(c-h)]. If the potential barrier is sufficiently large, the penetration depth may be characterized by $d \sim \frac{\hbar}{2\sqrt{2m(V_0-E)}}$, under the WKB approximation. It is evident that as the energy increases, the penetration depth also becomes larger. The ground-state wave function exhibits even parity, while the first excited state shows odd parity, and the second excited state exhibits even parity, and so forth: An even-odd-even-odd alternating parity pattern of QBS wave functions present. Based on the tunneling spectra (Fig. \ref{fig:ee_eis_trs}), the incident energies corresponding to RT, or the RT levels are readily obtained.    

\begin{figure*}[htp]
	\centering
	\includegraphics[width=0.95\linewidth]{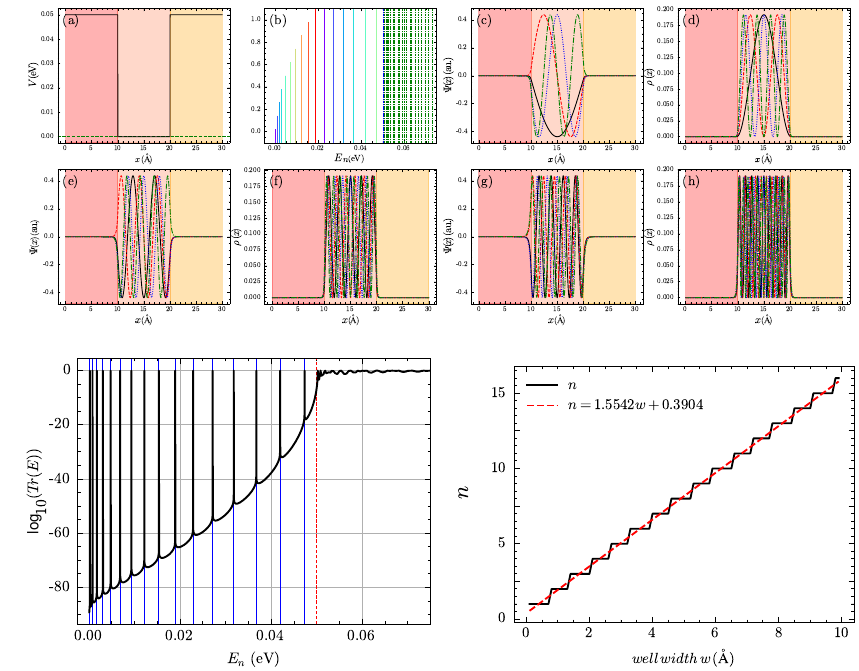}
	\caption{Similar to Fig. \ref{fig:ee_eis_trs} but for H atoms. The parameters for the double-barriers system are: $V_0$ = 0.05 $\rm{eV}$, $a = 10$\, \AA, $w = 10$\, \AA. The red dashed line of bottom panel represents the linear fit $n = 1.5542 w + 0.3904$. }
	\label{fig:H_eis_trs}
\end{figure*}

Furthermore, the full-width-at-half-maximum (FWHM) of each RT peak (i.e., the energy broadening, denoted by $\sigma$) can be deduced and used to estimate the lifetime of RT levels and consequently the lifetime of QBS levels. The calculated energy levels using the two methods are listed in Table \ref{tab:ee_eis_trs}, along with the energy broadening and parities of wave functions.  The DB system contains 15 QBS of electrons with the following parameters: barrier height $V_0 = 3\, $eV, barrier width $a = 10$ \AA, and well width $w = 50$ \AA. The results regarding the quantum nature of H atoms in a DB system, i.e., the QBS levels and wave functions, and quantum tunneling as a quantum particle, are depicted in Fig. \ref{fig:H_eis_trs} and compared in Table \ref{tab:H_eis_trs}. By examining the characteristics of wave functions displayed in Figs. \ref{fig:ee_eis_trs} and \ref{fig:H_eis_trs}, the following facts are evidenced: There is no node for the ground-state wave function, there is one node for the first excited state, and more generally, the nth QBS wave function $\psi_n$ has $(n-1)$ nodes. Such an observation is in line with Sturm’s theorem. 

As shown in right bottom part of both Figs. \ref{fig:ee_eis_trs} and \ref{fig:H_eis_trs}, the number of QBS levels shows a stepwise increase with the width of the potential well $w$, which can be approximated by the formula $n = A w + B$, a relationship that is inherently determined by the de Broglie wavelength $\lambda_d$ of the particle. In the case of an ideal infinite square potential well, stationary wave solutions are possible only when the width of the well is an integer or half-integer multiple of the de Broglie wavelength for a given particle energy. Likewise, in the scenario involving a double potential barrier, the condition $w_{n} \propto n\lambda_d$ is still satisfied \cite{yang2024Penetration}. 

From the precise RT levels and the QBS levels enumerated in Table \ref{tab:ee_eis_trs} for electrons and Table \ref{tab:H_eis_trs} for H atoms, it is clearly seen that the QBS levels obtained by exact diagonalization correspond one-to-one to the RT levels in the transmission spectrum, numerically confirming the aforementioned physical picture directly. On the other hand, due to the large difference of particle mass ($m_{e}/m_{H} \sim 1/1837$), the energy scale to show remarkable quantum effects at similar spatial scale is different. This is evidenced from the barrier height of DB systems under investigation: $V_0 = 3.0\, $ eV vs $V_0 = 0.05\, $ eV. The difference is understandable from the de Broglie wavelength $\lambda_d=\frac{h}{\sqrt{ 2m E}}$, which requires that the energy ratio $E_{H}/E_{e} = m_{e}/m_{H} \sim 1/1837$ for the same $\lambda_d$. Such a magnitude of barrier height can be encountered in the diffusion of H atoms on some realistic systems such as Pt(111) surface \cite{bi2021Atomic} or the graphene surface \cite{gonzalez2019Hydrogen} where the van der Waals interactions are dominant.

\renewcommand{\arraystretch}{1.5}
\begin{table*}
	\caption{Correspondence of the RT energies and the QBS
		levels of electrons. $\sigma$: energy broadening of RT peaks (equals to FWHM). The parities
		of the wave functions of the QBS P$(\psi_n)$ are listed. The parameters for the
		DB system are the same as in Fig. \ref{fig:ee_eis_trs}. }
	\label{tab:ee_eis_trs}
	\centering
	\resizebox{1.0\linewidth}{!}{
	\begin{tabular}{|>{\centering\arraybackslash}p{1cm}|>{\centering\arraybackslash}p{5cm}|>{\centering\arraybackslash}p{5cm}|>{\centering\arraybackslash}p{3cm}|>{\centering\arraybackslash}p{2cm}|}
			\hline
			$n$ & RT \,($\rm{eV}$)   & $\sigma$ \,($\rm{eV}$) & QB \,($\rm{eV}$) & P($\psi_{n}$) \\
			\hline
			1  & 0.01386653 & 7.29270116$\times 10^{-12}$  & 0.01376915    & even     \\
			\hline
			2  & 0.05545493 & 6.32678284$\times 10^{-11}$  & 0.05506559    & odd      \\
			\hline
			3  & 0.12473090 & 2.86959844$\times 10^{-10}$  & 0.12385562    & even     \\
			\hline
			4  & 0.22163489 & 7.84563775$\times 10^{-10}$  & 0.22008077    & odd      \\
			\hline
			5  & 0.34607818  & 1.79778014$\times 10^{-9}$ & 0.34365395    & even     \\
			\hline
			6  & 0.49793645  & 4.77584527$\times 10^{-9}$  & 0.49445318    & odd      \\
			\hline
			7  & 0.67703949  & 1.54043764$\times 10^{-8}$ & 0.67231165    & even     \\
			\hline
			8  & 0.88315491  & 4.62366565$\times 10^{-8}$  & 0.87700186    & odd      \\
			\hline
			9  & 1.11596083  & 1.32685825$\times 10^{-7}$  & 1.10820927    & even     \\
			\hline
			10 & 1.37499794  & 3.47897480$\times 10^{-7}$ & 1.36548613    & odd      \\
			\hline
			11 & 1.65957780  & 1.27538303$\times 10^{-6}$ & 1.64816373    & even     \\
			\hline
			12 & 1.96858512  & 5.41751317$\times 10^{-6}$  & 1.95516505    & odd      \\
			\hline
			13 & 2.29997023  & 2.62593281$\times 10^{-5}$  & 2.28453108    & even     \\
			\hline
			14 & 2.64903860  & 2.33287659$\times 10^{-4}$ & 2.63186215    & odd      \\
			\hline
			15 & 2.99971227  & 2.71227891$\times 10^{-3}$   & 2.98244627    & even     \\
			\hline
		\end{tabular}
	}
\end{table*}

\begin{table*}

	\caption{Similar to Table \ref{tab:ee_eis_trs} but for H atoms, with $V_0$ = 0.05 $\rm{eV}$,  $a = 10$ \AA,  $w = 10$ \AA.}
	\label{tab:H_eis_trs}
	\centering
	\resizebox{1.0\linewidth}{!}{

		\begin{tabular}{|>{\centering\arraybackslash}p{1cm}|>{\centering\arraybackslash}p{5cm}|>{\centering\arraybackslash}p{5cm}|>{\centering\arraybackslash}p{3cm}|>{\centering\arraybackslash}p{2cm}|}

			\hline
			$n$ & RT \,($\rm{eV}$)    & $\sigma$ \,($\rm{eV}$) & QB \,($\rm{eV}$) & P($\psi_{n}$) \\
			\hline
			1  & 0.00019028 & 1.29597178$\times 10^{-48}$ & 0.00018891    & even     \\
			\hline
			2  & 0.00076102 & 1.98128875$\times 10^{-47}$  & 0.00075554    & odd      \\
			\hline
			3  & 0.00171186  & 1.67422238$\times 10^{-46}$ & 0.00169953    & even     \\
			\hline
			4  & 0.00304220 & 1.26036784$\times 10^{-45}$ & 0.00302030    & odd      \\
			\hline
			5  & 0.00475115  & 1.70737103$\times 10^{-44}$ & 0.00471696    & even     \\
			\hline
			6  & 0.00683749  & 2.26009733$\times 10^{-43}$ & 0.00678833    & odd      \\
			\hline
			7  & 0.00929958  & 5.64252190$\times 10^{-42}$  & 0.00923278    & even     \\
			\hline
			8  & 0.01213524  & 1.30591663$\times 10^{-40}$ & 0.01204818    & odd      \\
			\hline
			9  & 0.01534157  & 7.63734659$\times 10^{-39}$  & 0.01523170    & even     \\
			\hline
			10 & 0.01891467  & 7.65211908$\times 10^{-37}$  & 0.01877952    & odd      \\
			\hline
			11 & 0.02284908  & 1.21086224$\times 10^{-34}$ & 0.02268632    & even     \\
			\hline
			12 & 0.02713692  & 5.70105862$\times 10^{-32}$  & 0.02694444    & odd      \\
			\hline
			13 & 0.03176596  & 6.25819418$\times 10^{-29}$  & 0.03154206    & even     \\
			\hline
			14 & 0.03671522  & 3.20658167$\times 10^{-25}$ & 0.03645906    & odd      \\
			\hline
			15 & 0.04194171  & 1.61833083$\times 10^{-20}$ & 0.04165478    & even     \\
			\hline
			16 & 0.04731380   & 1.03883988$\times 10^{-13}$ & 0.04700944    & odd      \\
			\hline
		\end{tabular}
	}
\end{table*}

Aside from the good agreement between the RT and QBS levels, there are still minor differences as seen from Table \ref{tab:ee_eis_trs} and \ref{tab:H_eis_trs} . The exact diagonalization method can in principle provide exact solutions for the system, including energy levels and wave functions which can be used to calculate the transfer probability between different states. Practically, high-precision numerical results from exact diagonalization require substantial computational resources. In the results presented in Fig. \ref{fig:ee_eis_trs}, the variation step size along the $x$-axis is 0.01 \AA, and the dimension of the matrix to be diagonalized is about $10000 \times 10000$. For simple potential barrier structures as that considered in this study, the TMM offers significantly higher computational efficiency for lower energy spacing. Nevertheless, to accurately determine the positions of resonant energy levels, it is necessary to employ specialized computational techniques to approximate the energy points where the transmission probability is 1. Taking the first data row in Table \ref{tab:ee_eis_trs} as an example, the FWHM of the lowest RT is $\sigma = 7.29270116 \times 10^{-12}$ eV which is very small compared to the energy levels of the system. This implies that the transmission probability drops very rapidly even a tiny deviation from the RT levels, hence a high energy resolution is required to accurately determine the exact numerical values of the RT levels. 

In our calculations, we have achieved a very high level of precision, by setting a convergence criterion of $|1 - Tr(E)| < 10^{-60}$, which strongly ensures that the system has experienced RT. The results of listed in Table \ref{tab:H_eis_trs} indicate that, compared to electrons, much more stringent condition has to be met with for the RT of H atoms, with the energy window (i.e., energy broadening $\sigma$) of the transmission spectrum being much narrower than in the case of electrons. For the lowest tunneling level ($E \sim 0.00019$ eV), H atoms can still completely traverse the double potential barriers, albeit requiring very precise incident energy with a broadening of $\sigma \sim 10^{-48}$ eV, comparing to the case of electron RT via the lowest QBS level ($\sigma \sim 10^{-12}$ eV). From both Table \ref{tab:ee_eis_trs} and \ref{tab:H_eis_trs}, it is also seen that the energy broadening for RT increases with higher levels, relaxing gradually the constraint on the monochromaticity of incident energy. Such a critical condition of RT provides an opportunity for measuring the energy of incident particles with ultrahigh accuracy.

\section{DEPENDENCE OF QUASI-BOUND STATES ON THE DOUBLE-BARRIER GEOMETRIES}\label{sec:wdith}

In this section a detailed analysis is carried out to study the dependence of QBS on the key parameters describing the double-barrier (DB) geometries: The inter-barrier spacing, the width of single barrier, and the barrier height.    
\subsection{ The Variation of QBS Levels with Inter-barrier Spacing}

It is evident that the QBS levels and/or RT levels are closely related to the inter-barrier spacing, that is, the quasi-well width. Before presenting the numerical results, we first conduct some theoretical analysis.
The transmission across a single rectangular barrier (the diagonal matrix element $m_{11}$ in Eq. (\ref{eq:transfer_matrix})) can be expressed as follows \cite{bi2021Quantum, bi2021Atomic, yang2024Penetration}
\begin{equation}\label{eq:m11}
	m_{11} = 2\gamma e^{-ika}[i(k^2-\beta^2)\sinh(\beta a) + 2k\beta\cosh(\beta a)],
\end{equation}
where $k = \sqrt{2mE/\hbar^{2}}$, $\beta = \sqrt{2m(V_{0} - E)/\hbar^{2}}$, and $\gamma = \frac{1}{4\beta k}$, $C_m=\sqrt{\frac{2m}{\hbar^2}}$. The Eq. (\ref{eq:m11}) can be rewritten as
\begin{eqnarray}
	m_{11}^2 &=&
	4\gamma^2\sigma^2 e^{i2(\alpha-C_m a\sqrt{E})},
\end{eqnarray}
where $\sigma = \sqrt{A^2 + B^2}$, $A = (k^2 - \beta^2)\sinh(\beta\alpha)$, $B = 2\beta k\cosh(\beta\alpha)$, and the angle $\alpha = \arctan( \frac{A}{B}) = \arctan(\delta\tanh(C_m a\sqrt{(V_{0} - E)}))$, $C_m \equiv \sqrt{\frac{2m}{\hbar^2}}$, $\delta \equiv (\frac{\beta}{k} - \frac{k}{\beta})$.
Drawing upon the theorem presented in Ref.  \cite{yang2024Penetration} the corresponding width of well is
$w = w_n = \frac{n\pi}{k}
	- \frac{\pi + \theta + 2ka}{2k}$, where $\theta = \arg(m_{11}^{2})$. The number of QBS levels $n$ scales stepwise with 
the width of the potential well $w$ with different step for a given incident energy $E$.

\begin{figure*}[htbp]
	\centering
	\includegraphics[width=0.95\linewidth]{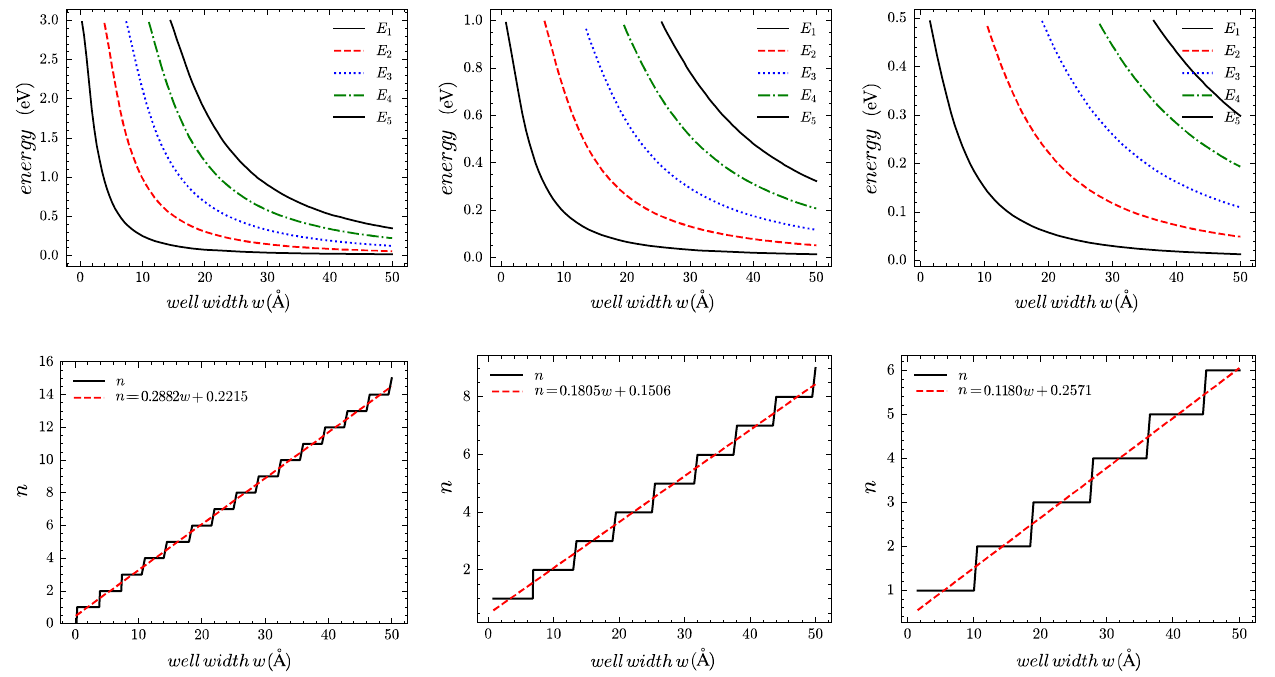}
	\caption{The dependence of the QBS on the inter-barriers spacing ($w$) at different barrier height $V_0 = 3.0,\, 1.0,\,0.5\,\rm{eV}$(from left to right) for electrons. Top panels: The first five energy levels ($E_1 $ to $E_5$). Bottom) panel: The number of QBS levels as a function of $w$. The red dashed line represents the linear fit $n = A w + B$.}
	\label{fig:ee_ei_n_w}
\end{figure*}

\begin{figure*}[hbtp]
	\centering
	\includegraphics[width=0.95\linewidth]{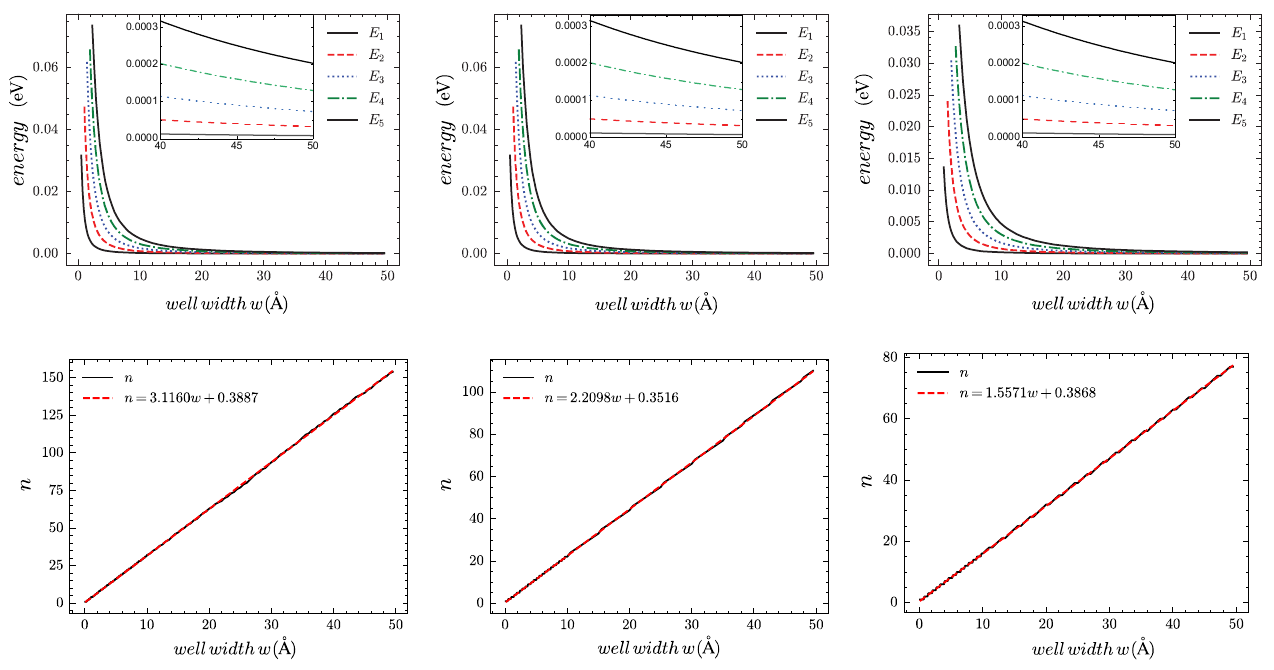}
	\caption{Similar to Fig. \ref{fig:ee_ei_n_w} but for H atoms at different barrier height $V_0 = 0.2,\, 0.1,\,0.05\,\rm{eV}$(from left to right).}
	\label{fig:H_ei_n_w}
\end{figure*}

The dependence of the QBS (energies and level counts $n$) on the inter-barriers spacing at different barrier heights for electrons and H atoms are shown in Fig. \ref{fig:ee_ei_n_w} and \ref{fig:H_ei_n_w}, respectively. Despite the different order of magnitudes, the QBS levels of electrons and H atoms exhibit a remarkably similar variation trend with inter-barriers spacing at various barrier heights.

\begin{figure*}[htp]
	\centering
	\includegraphics[width=0.95\linewidth]{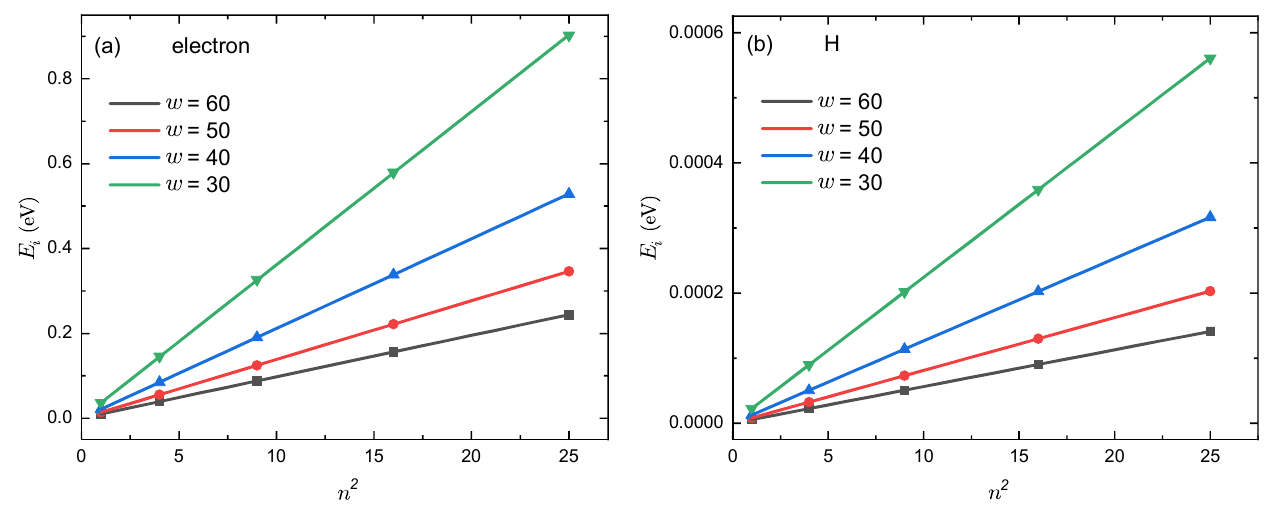}
	\caption{Dependence of QBS levels of electrons (left) and H atoms (right) on the square of the principle quantum number  $n$, for a number of quasi-well width $w$ (in units of \AA). }
	\label{fig:ni_vs_w}
\end{figure*}

When the term $ \frac{\pi + \theta +2ka}{2k}$ is negligible with comparison to the inter-barrier spacing (quasi-well width $w$), i.e.,  $ \frac{\pi + \theta +2ka}{2k}\ll w_n=w$, one approximately has $w\approx \frac{n\pi}{k}$, which yields $E \approx n^2 \frac{\pi^2\hbar^2}{2mw^2} $. This indicates that the QBS levels decrease monotonically with $w$, reducing the energy gap between each of the QBS, and eventually converge to the exact bound levels in an infinite-depth square potential well where $E_n = n^2 \frac{\pi^2\hbar^2}{2mw^2} $. This is independent of the geometries of the potential barriers. The variation trend is clearly demonstrated in Fig. \ref{fig:ni_vs_w}, where the first five QBS levels ($E_i$) of electron and H atom are shown along with the linearly fitted data lines as a function of the integer $n^2$.
Furthermore, compared to the RT levels across a finite-depth square  potential well, such a mathematical expression differs only by a constant (i.e., the well depth $V_0$) \cite{zeng2007Quantum}. The asymptotic coincidence of the QBS levels in arbitrarily shaped DBs with the exact solutions for an infinite-depth square potential well, and its similarity to the RT levels for a finite-depth square potential well reveal the intrinsic connection between these quantum systems.

The range of wave vectors for incident particles differs with varying barrier heights, leading to a variation in the number of energy levels and the linear coefficient associated with the quasi-well width, which is directly proportional to the maximum value of QBS energy ($\sqrt{E_{max}} \sim \sqrt{V_0} $ ). Consequently, the width of the energy level steps is inversely proportional to height of barrier, as evidenced from Fig. \ref{fig:ee_ei_n_w}. As pointed out in Ref. \cite{yang2024Penetration}, the resonant states (equivalently, the QBS) appears periodically with inter-barrier spacing (quasi-well width) $w$ via the variation step $\Delta w=\frac{\pi}{k}\sim \frac{h}{2\sqrt{2mV_0}}$, which is just the step width associated with the variation of QBS numbers. In the case of H atoms, the much larger particle mass and consequently the much smaller $\Delta w$ leads to an almost linear variation of QBS numbers with $w$ (see lower panels of Fig. \ref{fig:H_ei_n_w}).

\subsection{Parameter Space for QBS}

\begin{figure*}[htp]
	\centering
	\includegraphics[width=0.95\linewidth]{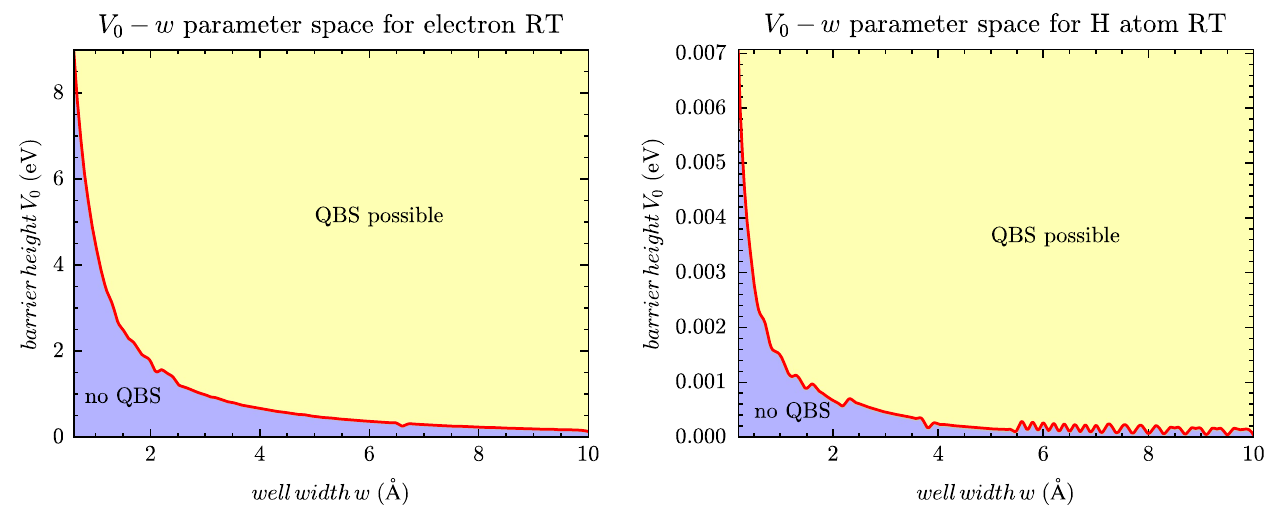}
	\caption{$V_0-w$ parameter space for the RT of electrons (left) and H atoms (right) at a given barrier width ($a = 10$ \AA).}
	\label{fig:phase}
\end{figure*}
Upon examining Figs. \ref{fig:ee_ei_n_w} and \ref{fig:H_ei_n_w}, it becomes evident that RT of quantum particles is not ubiquitous across varying barrier heights and quasi-well widths. Specifically, it can be expected that QBS are absent when the quasi-well is narrow enough or the barrier height is not large enough. Such a constraint is intrinsically contained in the mathematical expression of $w_n$, which gives that $w_1=\frac{(\pi-\theta)\hbar}{2\sqrt{2mV_0}}-a$. It requires that $w \geq w_1$ to have RT to take place in a DB system, or equivalently, to guarantee at least one QBS presents in the quasi-well region. The necessary and sufficient condition to have only one QBS level is therefore $w_1 \leq w \leq \Delta w  =\frac{\pi}{k}$.  The results of more generalized analysis are shown in Fig. \ref{fig:phase}, where a boundary line can be drawn in the parameter space of the barrier height ($V_0$) versus well width ($w$), for the existence/absence of QBS of electrons and H atoms. It is evident that RT is possible only when the system’s parameters are positioned above the boundary line. On the boundary, there exists exactly one QBS, and its energy level is close to the barrier height. Utilizing this diagram, we are able to ascertain the specific parameters related to DB systems that allow the occurrence of RT. The $V_0-w$ parameter constraint on the existence of QBS levels is significantly different from the case of a finite-depth square potential well, where at least one even-parity bound state exists in spite of the width and depth of the potential well \cite{zeng2007Quantum}.

\begin{figure}
	\centering
	\includegraphics[width=0.95\linewidth]{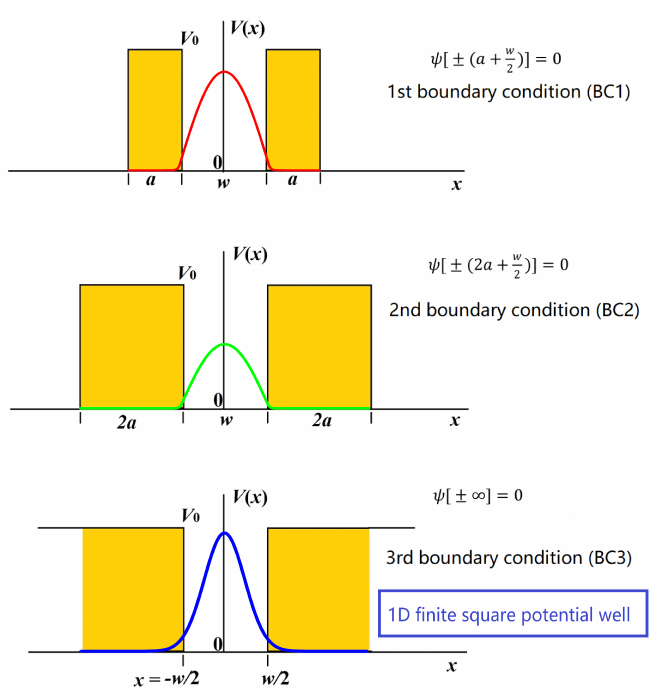}
	\caption{Schematic diagrams for different boundary conditions of DB systems, where varying barrier widths present.}
	\label{fig:compare_ei}
\end{figure}

\renewcommand{\arraystretch}{1.3}
\begin{table*}

	\caption{The QBS levels of electrons in rectangular DB, calculated using different types of boundary conditions (BC) as schematically shown in Fig. \ref{fig:compare_ei}: $\psi[\pm(a+\frac{w}{2})]=0$ for BC1, $\psi[\pm(2a+\frac{w}{2})]=0$ for BC2 and $\psi[\pm\infty]=0$ for BC3. The parameters describing the DB system: $V_0$ = 3.0 $\rm{eV}$, $w = 50\,$ \AA.}
	\label{tab:compare_ei}
	\centering
	\resizebox{1.0\linewidth}{!}{
		\begin{tabular}{|>{\centering\arraybackslash}p{1cm}|>{\centering\arraybackslash}p{5cm}|>{\centering\arraybackslash}p{5cm}|>{\centering\arraybackslash}p{3cm}|}
			\hline
			$n$ & $E_{BC1} \,(\rm{eV}$) & $E_{BC2} \,(\rm{eV}$) & $E_{BC3} \,(\rm{eV}$) \\		
			\hline
			1 & 0.013768 & 0.013766 & 0.013771 \\
			\hline
			2 & 0.055064 & 0.055052 & 0.055071 \\
			\hline
			3 & 0.123852 & 0.123826 & 0.123868 \\
			\hline
			4 & 0.220074 & 0.220027 & 0.220103 \\
			\hline
			5 & 0.343644 & 0.343570 & 0.343697 \\
			\hline
			6 & 0.494438 & 0.494332 & 0.494548 \\
			\hline
			7 & 0.672291 & 0.672147 & 0.672531 \\
			\hline
			8 & 0.876975 & 0.876787 & 0.877498 \\
			\hline
			9 & 1.108175 & 1.107937 & 1.109278 \\
			\hline
			10 & 1.365443 & 1.365151 & 1.367682 \\
			\hline
			11 & 1.648110 & 1.647758 & 1.652489 \\
			\hline
			12 & 1.955096 & 1.954681 & 1.963466 \\
			\hline
			13 & 2.284425 & 2.283923 & 2.300354 \\
			\hline
			14 & 2.631483 & 2.632941 & 2.662871 \\
			\hline
			15 & 2.973093 & 2.983424 & 3.050714 \\
			\hline
		\end{tabular}
	}
\end{table*}

\subsection{Effects of Boundary Conditions on QBS Levels}
In one-dimensional systems, the Schrödinger equation is an ordinary differential equation for which the boundary conditions play a nontrivial role. In this subsection, we study how the boundary conditions and double-barrier geometries would affect the eigenvalues (QBS levels). Specially, we study the dependence of QBS levels on barrier width, for a symmetrical DB with a fixed inter-barrier spacing (quasi-well width) $w$. As illustrated in Fig. \ref{fig:compare_ei}, we pay attention to the investigation of three types of boundary conditions and delve into the effects on the QBS levels of the DB systems. Table \ref{tab:compare_ei} summarizes the QBS levels obtained by the exact diagonalization at different DB boundary conditions. Despite the small magnitude of variations, the impact of different barrier widths on the position of QBS energy levels is generally negligible.

\section{Tunable Optical Properties of the Double-Barrier systems}\label{sec:opab}

In this section, we study the light absorption properties of the system. The response functions, i.e., the dielectric function $\epsilon = \epsilon_{1} + i \epsilon_{2}$ and the optical conductivity $\sigma = \sigma_{1} + i \sigma_{2}$, are pivotal in characterizing the interactions between applied electromagnetic fields and materials. These functions encompass both real and imaginary components, which are crucial for understanding the material's response to electromagnetic waves.
In general, $\epsilon(\omega) $ and $\sigma(\omega) $, represent complex-valued functions of angular frequency $\omega$. The real component of the optical conductivity, denoted as $\sigma_{1}$, is instrumental in determining the absorption within the medium, as it influences the imaginary part of the dielectric function, $\epsilon_{2}$. Conversely, the imaginary component of the optical conductivity, $\sigma_{2}$, contributes to the real part of the dielectric function, $\epsilon_{1}$, which in turn affects the polarization of the material \cite{dresselhaus2001SOLID}. The components $\epsilon_{1}$ and $\epsilon_{2}$ are critical for understanding how the material influences the propagation of light. Specifically, $\epsilon_{1}$ provides insights into the retardation of light's velocity, while $\epsilon_{2}$ accounts for the absorption and loss of light energy due to polarization as it traverses across the material medium. A comprehensive understanding of the dielectric function is vital for the analysis and application of materials in various optical and electronic devices.

For simplicity without losing generality of the results, we consider the situation in which the incident light wave is a monochromatic plane wave, to investigate the optical properties of DB systems based on electric dipole transition. From the picture of medium absorption \cite{huang1998Solid}, the imaginary part of the corresponding dielectric function $\epsilon_{2}(\omega)$ can be given as follows (refer to the details in Appendix \ref{app:epsilon})
\begin{widetext}
\begin{eqnarray}\label{eq:epsilon_2}
	\epsilon_{2}(\omega) &=& \frac{\pi e^2}{6\hbar \epsilon_{0} \Omega_{0}}
	|\langle k|r|n\rangle|^2J_{nk}(\omega)
	 = \frac{\pi e^2}{6\hbar \epsilon_{0} w}\frac{|\langle k|r|n\rangle|^2}{S_0}\delta(E_n - E_k - \hbar \omega)
\end{eqnarray}
\end{widetext}
where $\epsilon_{0}$ is the vacuum permittivity, $e$ is elementary electric charge, $\Omega_0$ is the volume of the well region, $w$ is the width of the barrier, $r$ is the electron coordinate in real space and $S_0$ is the cross section area of the quasi-well region within the DB system. $J_{nk}(\omega) = \delta(E_n - E_k - \hbar \omega)$ is the joint density of states (JDOS), representing the energy level distribution corresponding to the quantum transition between the $n$th and $k$th level with an energy difference of $E_{nk}=\hbar \omega $.In semiconducting systems, it corresponds to the JDOS as determined by the valence and conduction bands  \cite{yu2010Fundamentals}. The term $|\langle k|r|n\rangle|^2$ determined by the energy level distribution and the transition matrix, reflects the constraints imposed by the selection rules on the light absorption spectra of the electron bound states.

\begin{figure*}[htbp]
	\centering
	\includegraphics[width=0.95\linewidth]{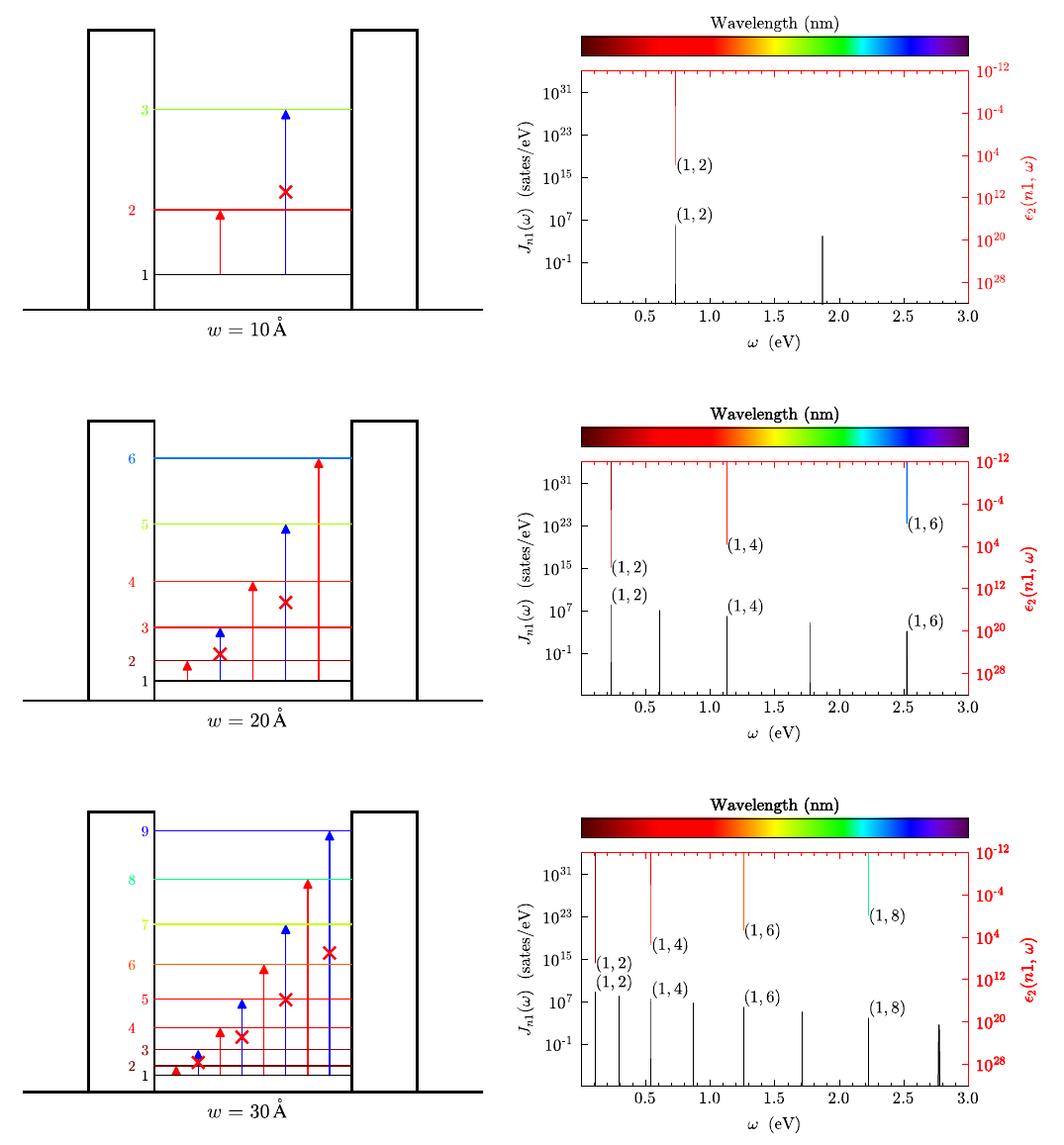}
	\caption{The optical absorption properties of DB system of electrons. Left panels: Selection rules for transitions between QBS energy levels $(k \rightarrow n)$. Red arrows denote permitted transitions, whereas blue arrows marked with a red cross indicate forbidden transitions. Right panels: The JDOS $J_{nk}(\omega)$ and the imaginary part of the dielectric function $\epsilon_{2,nk}(\omega)$ corresponding to transitions between different energy levels $(k \rightarrow n)$. The absorption of photon energy $\hbar \omega = E_{nk} = E_{n} - E_{k}$, which corresponds to the energy level differences, indicated by wavelength $\lambda \approx 1240/\Delta E$ (nm), is represented by different colors from red to purple. The color scale corresponds to an energy range of [0.0\,-\,3.0]eV or a wavelength range [$\infty$\, - 413.3] nm. The parameters same as in Table \ref{tab:opab_11}.}
	\label{fig:opab_11}
\end{figure*}

Utilizing the wave functions obtained from extract diagonalization, we have directly computed the matrix elements $|\langle k|r|n\rangle|^2$ and the complex dielectric function $\epsilon_{2}$ numerically. In the calculations, we take $ k = 1,\,2,\,3 $ (occupied states), $ n = 2,\,3,\,\ldots,\,N $ (unoccupied states); $ N $ represents the total number of QBS levels. For the quasi-well width $ w = 10,\,20,\, 30, \,40, \,50,\, 60 \,$ \AA, with a cross-sectional area $ S_0 = a^2 = 100\, $ \AA$^2$, the quantities  $ J_{nk}(\omega) $ and $ \epsilon_{2,nk}(\omega) $ have been calculated separately. This approach will provide a direct insight into how the spacing of the potential barriers influences the distribution of electronic energy levels and, consequently, the modulation of optical absorption characteristics. The numerical results are presented in Tables \ref{tab:opab_11}-\ref{tab:opab_12} and Tables \ref{tab:opab_21}-\ref{tab:opab_32} ( see Appendix \ref{app:table}) and Figs. \ref{fig:opab_11}-\ref{fig:opab_12}.

\begin{figure*}[htbp]
	\centering
	\includegraphics[width=0.95\linewidth]{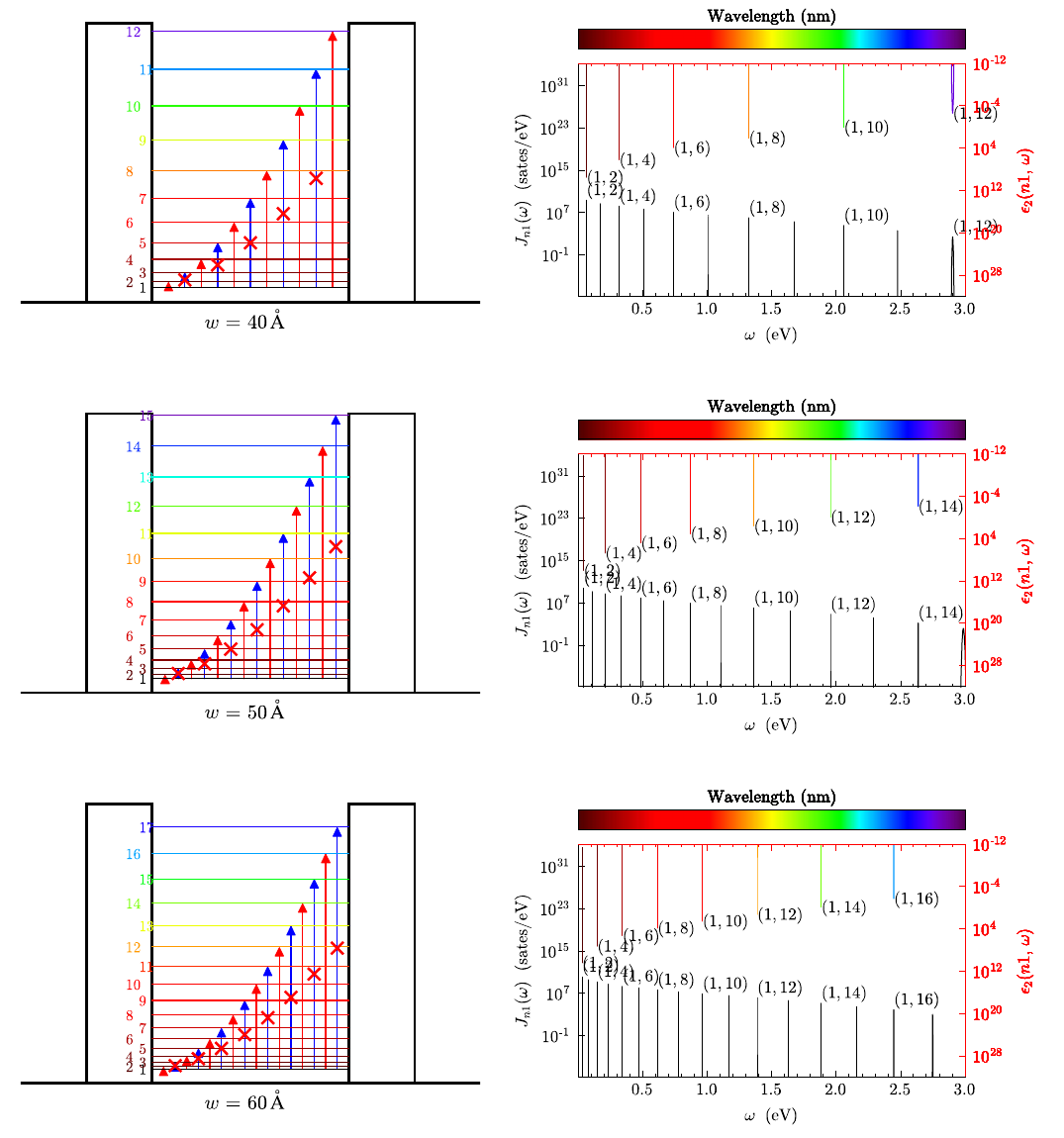}
	\caption{The optical absorption properties of DB system of electrons at different well widths $w = 40, \,50, \,60$ \AA, similar to Fig. \ref{fig:opab_11}.}
	\label{fig:opab_12}
\end{figure*}

First, we study the characteristics of JDOS and show the effects of optical selection rules. In the context of electric dipole transitions, the electric dipole operator $ \vec{D} = -e \vec{r} $ is an odd function of coordinates, therefore non-zero matrix elements $|\langle k|r|n\rangle|^2 \neq 0$ is possible only for transitions between the states with opposite parity of wave functions. The corresponding selection rule is therefore
\begin{equation}\label{eq:select}
	\text{Odd parity state}\, \leftrightarrows \,\text{Even parity state}
\end{equation}
For the one-dimensional systems discussed herein, the wave functions are plane waves along the $x$ axis, it is only necessary to consider the parity selection rule. Additionally, the matrix element representing the transition probability is largely determined by the overlap of the wave functions of the initial and final states. A larger difference in principal quantum numbers can lead to a more significant overlap, especially in cases where the electron hops to a more diffuse orbital (see the wave functions in Fig. \ref{fig:ee_eis_trs}).
Our numerical data corroborate this assertion. Taking results from Table \ref{tab:opab_11} as an example, we calculated the optical absorption properties for transitions from the $ k=1 $ state to the $ n $th states when the well width $w = 10,\, 20\, \rm{and}\, 30$ \AA. The results demonstrate that effective absorption only occurs between the $ k=1 $ state and states whose quantum number $n$ is an even number, where a non-zero $ \epsilon_2$ presents. It is evident that this is determined by the parity of the wave functions of QBS, adhering to the parity selection rule. As presented in Fig. \ref{fig:ee_eis_trs} and Table \ref{tab:ee_eis_trs}, the principle quantum number $n$ designating the QBS levels exhibits consistent parity characteristics with the wave functions.

\renewcommand{\arraystretch}{1.5}
\begin{table*}
	\caption{The optical absorption properties of DB system of electrons at different well widths $w = 10, \,20, \,30$ \AA. For each well width, from left to right, they are as follows: The absorption of photon energy $\hbar \omega = E_{nk} = E_{n} - E_{k}$, the FWHM of resonant peak $\Gamma_{n}$, the JDOS $J_{nk}(\omega)$ and the imaginary part of the dielectric function $\epsilon_{2,nk}(\omega)$ corresponding to transitions between different energy levels $(k = 1 \rightarrow n)$. The symbol "---" indicates forbidden transitions. The units of $\hbar \omega$ and $\Gamma_{n}$ are eV and the unit of $J_{nk}(\omega)$ is $\rm{eV}^{-1}$. Parameters : the barrier height $V_0 = 3.0$ eV, the width of the barrier $a = 10$ \AA, the inter-barrier potential well width (from top to bottom) $w = 10,\,20,\,30$ \AA, the initial QBS number $k = 1$. }
	\label{tab:opab_11}
	\centering
	\resizebox{1.0\linewidth}{!}{

		\begin{tabular}{|c|c|c|c|c|c|c|c|c|c|c|c|c|}

			\hline
			$w$  & \multicolumn{4}{c|}{10} & \multicolumn{4}{c|}{20} & \multicolumn{4}{c|}{30}                                                                    \\
			\hline
			$n$ & $\hbar \omega$     & $\Gamma_{n}$      & $J_{nk}$        & $\epsilon_2$ & $\hbar \omega$ & $\Gamma_{n}$ & $J_{nk}$  & $\epsilon_2$ & $\hbar \omega$ & $\Gamma_{n}$ & $J_{nk}$  & $\epsilon_2$ \\
			\hline
			2  & 0.7334         & 3.1721$\times 10^{-7}$       & 1.2577$\times 10^{6}$       & 6.0601$\times 10^{5}$  & 0.2282     & 3.5565$\times 10^{-9}$  & 1.1217$\times 10^{8}$ & 8.5992$\times 10^{7}$  & 0.1090     & 6.2676$\times 10^{-10}$  & 6.3652$\times 10^{8}$ & 6.8027$\times 10^{8}$  \\
			\hline
			3  & 1.8676         & 4.4571$\times 10^{-5}$       & 8.9507$\times 10^{3}$       & ---     & 0.6056     & 3.4509$\times 10^{-8}$  & 1.1560$\times 10^{7}$ & ---     & 0.2903     & 3.3568$\times 10^{-9}$  & 1.1884$\times 10^{8}$ & ---     \\
			\hline
			4  &             &             &             &       & 1.1261     & 4.3497$\times 10^{-7}$  & 9.1718$\times 10^{5}$ & 4.5738$\times 10^{3}$  & 0.5430     & 1.2890$\times 10^{-8}$  & 3.0949$\times 10^{7}$ & 2.1280$\times 10^{5}$  \\
			\hline
			5  &             &             &             &       & 1.7771     & 7.8591$\times 10^{-6}$  & 5.0762$\times 10^{4}$ & ---     & 0.8660     & 6.9619$\times 10^{-8}$  & 5.7304$\times 10^{6}$ & ---     \\
			\hline
			6  &             &             &             &       & 2.5225     & 2.9148$\times 10^{-4}$  & 1.3687$\times 10^{3}$ & 5.1509$\times 10^{-1}$  & 1.2571     & 4.4837$\times 10^{-7}$  & 8.8977$\times 10^{5}$ & 4.6845$\times 10^{2}$  \\
			\hline
			7  &             &             &             &       &        &       &      &       & 1.7128     & 3.3706$\times 10^{-6}$  & 1.1836$\times 10^{5}$ & ---     \\
			\hline
			8  &             &             &             &       &        &       &      &       & 2.2251     & 4.6616$\times 10^{-5}$  & 8.5580$\times 10^{3}$ & 7.6659$\times 10^{-1}$  \\
			\hline
			9  &             &             &             &       &        &       &      &       & 2.7703     & 9.3803$\times 10^{-4}$  & 4.2530$\times 10^{2}$ & ---     \\
			\hline
		\end{tabular}
	}
\end{table*}

\begin{table*}
	\caption{The optical absorption properties of DB system of electrons at different well widths $w = 40, \,50, \,60$ \AA \, and $k = 1$, similar to Table \ref{tab:opab_11}.}
	\label{tab:opab_12}
	\centering
	\resizebox{1.0\linewidth}{!}{

		\begin{tabular}{|c|c|c|c|c|c|c|c|c|c|c|c|c|}

			\hline
			$w$  & \multicolumn{4}{c|}{40} & \multicolumn{4}{c|}{50} & \multicolumn{4}{c|}{60}                                                                    \\
			\hline
			$n$ & $\hbar \omega$     & $\Gamma_{n}$      & $J_{nk}$        & $\epsilon_2$ & $\hbar \omega$ & $\Gamma_{n}$ & $J_{nk}$  & $\epsilon_2$ & $\hbar \omega$ & $\Gamma_{n}$ & $J_{nk}$  & $\epsilon_2$ \\
			\hline
			2  & 0.0636         & 1.8253$\times 10^{-10}$       & 2.1856$\times 10^{9}$       & 3.0029$\times 10^{9}$  & 0.0416     & 6.3268$\times 10^{-11}$  & 6.3056$\times 10^{9}$ & 1.0595$\times 10^{10}$  & 0.0293     & 3.1975$\times 10^{-11}$  & 1.2477$\times 10^{10}$ & 2.4790$\times 10^{10}$  \\
			\hline
			3  & 0.1694         & 8.2558$\times 10^{-10}$       & 4.8323$\times 10^{8}$       & ---     & 0.1109     & 2.8696$\times 10^{-10}$  & 1.3902$\times 10^{9}$ & ---     & 0.0781     & 1.0581$\times 10^{-10}$  & 3.7703$\times 10^{9}$ & ---     \\
			\hline
			4  & 0.3174         & 2.3578$\times 10^{-9}$       & 1.6920$\times 10^{8}$       & 1.4913$\times 10^{6}$  & 0.2078     & 7.8456$\times 10^{-10}$  & 5.0849$\times 10^{8}$ & 5.4746$\times 10^{6}$  & 0.1465     & 2.7503$\times 10^{-10}$  & 1.4505$\times 10^{9}$ & 1.8459$\times 10^{7}$  \\
			\hline
			5  & 0.5071         & 8.4105$\times 10^{-9}$       & 4.7434$\times 10^{7}$       & ---     & 0.3322     & 1.7978$\times 10^{-9}$  & 2.2191$\times 10^{8}$ & ---     & 0.2342     & 7.2055$\times 10^{-10}$  & 5.5366$\times 10^{8}$ & ---     \\
			\hline
			6  & 0.7383         & 3.3829$\times 10^{-8}$       & 1.1793$\times 10^{7}$       & 7.9204$\times 10^{3}$  & 0.4841     & 4.7758$\times 10^{-9}$  & 8.3533$\times 10^{7}$ & 6.8405$\times 10^{4}$  & 0.3414     & 2.0686$\times 10^{-9}$  & 1.9286$\times 10^{8}$ & 1.8651$\times 10^{5}$  \\
			\hline
			7  & 1.0103         & 1.2283$\times 10^{-7}$       & 3.2479$\times 10^{6}$       & ---     & 0.6632     & 1.5404$\times 10^{-8}$  & 2.5898$\times 10^{7}$ & ---     & 0.4680     & 3.8962$\times 10^{-9}$  & 1.0239$\times 10^{8}$ & ---     \\
			\hline
			8  & 1.3222         & 3.7955$\times 10^{-7}$       & 1.0511$\times 10^{6}$       & 1.2019$\times 10^{2}$  & 0.8693     & 4.6237$\times 10^{-8}$  & 8.6283$\times 10^{6}$ & 1.2000$\times 10^{3}$  & 0.6138     & 9.4192$\times 10^{-9}$  & 4.2354$\times 10^{7}$ & 6.9485$\times 10^{3}$  \\
			\hline
			9  & 1.6726         & 2.0394$\times 10^{-6}$       & 1.9562$\times 10^{5}$       & ---     & 1.1021     & 1.3269$\times 10^{-7}$  & 3.0067$\times 10^{6}$ & ---     & 0.7788     & 2.0561$\times 10^{-8}$  & 1.9403$\times 10^{7}$ & ---     \\
			\hline
			10 & 2.0588         & 1.0996$\times 10^{-5}$       & 3.6279$\times 10^{4}$       & 1.0670$\times 10^{0}$  & 1.3611     & 3.4790$\times 10^{-7}$  & 1.1467$\times 10^{6}$ & 4.1012$\times 10^{1}$  & 0.9628     & 5.0196$\times 10^{-8}$  & 7.9476$\times 10^{6}$ & 3.3483$\times 10^{2}$  \\
			\hline
			11 & 2.4751         & 1.0379$\times 10^{-4}$       & 3.8439$\times 10^{3}$       & ---     & 1.6457     & 1.2754$\times 10^{-6}$  & 3.1280$\times 10^{5}$ & ---     & 1.1656     & 1.1066$\times 10^{-7}$  & 3.6050$\times 10^{6}$ & ---     \\
			\hline
			12 & 2.9027         & 1.4433$\times 10^{-3}$       & 2.7642$\times 10^{2}$       & 2.5154$\times 10^{-3}$  & 1.9547     & 5.4175$\times 10^{-6}$  & 7.3639$\times 10^{4}$ & 8.7362$\times 10^{-1}$  & 1.3870     & 2.9400$\times 10^{-7}$  & 1.3570$\times 10^{6}$ & 1.8959$\times 10^{1}$  \\
			\hline
			13 &             &             &             &       & 2.2861     & 2.6259$\times 10^{-5}$  & 1.5192$\times 10^{4}$ & ---     & 1.6266     & 9.1721$\times 10^{-7}$  & 4.3495$\times 10^{5}$ & ---     \\
			\hline
			14 &             &             &             &       & 2.6352     & 2.3329$\times 10^{-4}$  & 1.7101$\times 10^{3}$ & 7.8839$\times 10^{-3}$  & 1.8838     & 2.9447$\times 10^{-6}$  & 1.3548$\times 10^{5}$ & 7.4663$\times 10^{-1}$  \\
			\hline
			15 &             &             &             &       & 2.9858     & 2.7123$\times 10^{-3}$  & 1.4709$\times 10^{2}$ & ---     & 2.1578     & 1.3265$\times 10^{-5}$  & 3.0076$\times 10^{4}$ & ---     \\
			\hline
			16 &             &             &             &       &        &       &      &       & 2.4467     & 5.1916$\times 10^{-5}$  & 7.6844$\times 10^{3}$ & 1.8845$\times 10^{-2}$  \\
			\hline
			17 &             &             &             &       &        &       &      &       & 2.7466     & 3.7245$\times 10^{-4}$  & 1.0711$\times 10^{3}$ & ---     \\
			\hline
		\end{tabular}
	}
\end{table*}

The time scale related to the light absorption process can be investigated through the JDOS $ J_{nk}(\omega) $. By approximating the delta function $\delta(E_n - E_k - \hbar \omega)$ with a normalized Gaussian function $ \frac{1}{\sqrt{2\pi}\Gamma_n}e^{-(\frac{E_n - E_k - \hbar \omega}{\Gamma_n})^2} $ and leveraging the energy-time uncertainty relationship, the inverse of broadening $ \Gamma_n $ reflects the lifetime of optical transition. The broadening of the lowest QBS energy level $\sigma_1$ is minimal and significantly smaller than that of higher energy levels (See Table \ref{tab:ee_eis_trs}). Therefore, it is reasonable to approximate the broadening of the JDOS $\Gamma_n$ with the broadening of the final state $\sigma_n$. The JDOS at $\hbar \omega = E_n - E_k = E_{nk}$ can be rewritten as $ J_{nk}(\omega) = \frac{1}{\sqrt{2\pi}\Gamma_n}$. Clearly, the level broadening $\sigma_n$ intensifies with the increment of $ n $, the peaks of the JDOS decrease rapidly, and the associated absorption lifetime reduces correspondingly. From Table \ref{tab:opab_11} and \ref{tab:opab_12}, the JDOS varies from $10^2$ to $10^{10}$ eV$^{-1}$. In natural units, $1 \text{ eV}^{-1} \approx 6.582 \text{ fs}$, hence the lifetime of the light absorption process spans the range of $\sim 0.1$ ps to 10 $\mu$s. This timescale is considerably longer than the typical value ($\sim $ ps) observed for optical transitions in semiconductors \cite{yu2010Fundamentals, trovatello2020Ultrafast, chen2023Recent}.

The imaginary part of the dielectric function $\epsilon_{2}$ for a number of transition processes are listed in Tables \ref{tab:opab_11} -\ref{tab:opab_32} (Tables \ref{tab:opab_21}-\ref{tab:opab_32}, see Appendix \ref{app:table}). Obviously, a larger difference in the principal quantum number $n$, corresponds to the absorption/emission of higher-energy (and typically shorter wavelength) photons, which is a general feature observed in the gaseous phase of atoms or molecules. This principle is employed in the design of solar cells and light-emitting diodes (LEDs), where the absorption and emission of light are critical. By adjusting the Fermi level, the DB systems based on real materials can be designed to absorb a specific range of the solar spectrum more efficiently or emit light at desired wavelengths. Considering the extraordinary light absorption capabilities and the unprecedented energy resolution of DB systems, it is possible to engineer systems that can accurately measure the energy of electromagnetic waves across the domain of  infrared/THz and even into the visible spectrum. With an energy resolution as low as $10^{-10}$ eV, these systems offer an unprecedented level of sensitivity and precision. Such refinement would be invaluable for a spectrum of scientific and technological applications, including sophisticated spectroscopic analysis, quantum optical experiments, and the innovation of cutting-edge optoelectronic devices \cite{jones2019Raman, huang2022Highresolution, munsi2024Energy}.

To elucidate the role of selection rules in the light absorption process and to showcase the influence of varying well widths more effectively, Figs. \ref{fig:opab_11} and \ref{fig:opab_12} depict schematic diagrams of the $k=1 \rightarrow n$ transitions, along with the JDOS and absorption spectra. The left panels employ red arrows to indicate permissible transitions, while blue arrows with red crosses symbolize the prohibited transitions, providing a clear visual representation of the selection rules as illustrated by Eq.(\ref{eq:select}). The absorption spectra on the right, which closely resembles the absorption spectrum of elements, with discrete spectral lines representing transitions from $k=1$ to various final states $n$. It is evident that the JDOS $J_{nk}$ on the left vertical axis always present for any $n$, while the imaginary part of the dielectric function $\epsilon_{2}$ on the right vertical axis exhibits spectral lines only for transitions that are allowed by the selection rules. To enhance the spectral line features, we have utilized a coloring scheme akin to that of elemental absorption spectra, with the hue of each line determined by its energy. 
As seen from Figs. \ref{fig:opab_11} and \ref{fig:opab_12}, and the data listed in Tables \ref{tab:opab_11}-\ref{tab:opab_32} (Tables \ref{tab:opab_21}-\ref{tab:opab_32}, see Appendix \ref{app:table}), the imaginary part of the dielectric function $\epsilon_{2,nk}(\omega)$ for the transition between the low-lying energy levels (e.g., $1\,\rightarrow\,2$, $1\,\rightarrow\,4$, $1\,\rightarrow\,6$, $2\,\rightarrow\,3$, $2\,\rightarrow\,5$) varies from the order of $10^2$ to $10^{10}$ for photon energies $\hbar\omega \sim 0.02$ to 0.7 eV, which is one to nine orders of magnitude higher than that of typical semiconductors \cite{yu2010Fundamentals, brown2021Electroluminescence}. For photons with an energy  $\hbar\omega=E_{nk}=E_n-E_k$ that excites effective transition from the  $k$th state to the  $n$th state with a sum energy broadening of  $\Delta E_{nk} \sim (\Gamma_n + \Gamma_k)$, the variation range of frequency ($\Delta \nu$) and wavelength ($\Delta \lambda$) is therefore $\frac{|\Delta \nu_{nk}|}{\nu_{nk}} = \frac{|\Delta \lambda_{nk}|}{\lambda_{nk}}= \frac{|\Delta E_{nk}|}{E_{nk}}$. In the case of $1 \longrightarrow 2$ transitions, the relative resolution for frequency and wavelength can be as high as $10^{-8}$ to $10^{-10}$ (see e.g., Table \ref{tab:opab_11}). Such intense optical absorption features point to the possibility of ultrahigh-precision detection of infrared or THz radiations based on nano-sized DB systems \cite{eaton2016Semiconductor}.     

\begin{figure*}[htp]
	\centering
	\includegraphics[width=0.95\linewidth]{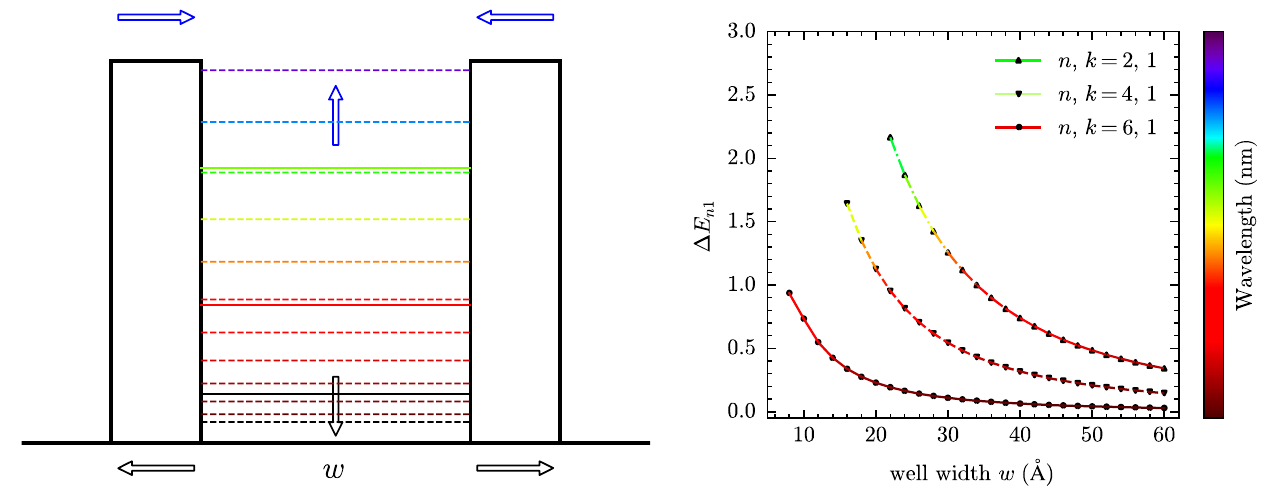}	
	\caption{Schematic diagram of the regulation of the optical properties of the well region by the inter-barrier spacing (quasi-well width). (left) The black arrow signifies that an increase in the well width leads to a reduction in the QBS levels; the blue arrow denotes that a decrease in the well width results in a higher energy of the absorbed light for transitions involving states with identical principal quantum numbers. (right) the energy difference between QBS energy levels with $k=1$ and $n= 2,\,4,\,6$ (from top to bottom). The color scale corresponds to a energy range of [0.0\,-\,3.0]eV, or a wavelength range [$\infty$\, - 413.3]nm . The height of the barrier $V_0 = 3.0$ eV. }
	\label{fig:opab_w}
\end{figure*}

Now we study the impact of varying quasi-well widths. As shown in Fig. \ref{fig:opab_w}, for a given ($k,n$), as the well width increases, an increase in the well width correlates with a decrease in the energy of the absorbed light (resulting in longer wavelengths), and the corresponding level broadening becomes smaller. Generally, for the optical transition between two states with principal quantum numbers $k$ and $n$ ($k < n$),  the resulted change in band gap  due to a small variation of quasi-well width 
$\Delta w$ may be estimated as follows: $\Delta E_g \sim - \frac{(n^2-k^2)\hbar^2\pi^2}{2mw^2}\times2(\frac{\Delta w}{w}) = -2E_g \frac{\Delta w}{w}$.
Therefore, a reduced quasi-well width leads to an enlarged band gap and vice versa. On the other hand, the  the JDOS increases with well width, which in turn enhances the strength of light absorption. The width of the potential barrier spacing (the quasi-well width),  plays a pivotal role in modulating the optical properties of the well region in a quantum system. By engineering a DB structure with tailored well widths and barrier heights, one can precisely control the energy of the absorbed light to fall within a specific range. For example, with a barrier height $V_0=3.0 $ eV and well width of $w = 24\, $ \AA, the absorption spectrum is tuned to fall into the spectrum of the visible light.

\section{Conclusions}

In this study, we have conducted an extensive investigation into the RT of electrons and hydrogen atoms in DB systems. Our numerical computations provide direct evidence for the one-to-one correspondence between the RT energies and QBS levels. Detailed analyses reveal how inter-barrier spacing modulates the emergence and proliferation of QBS, as well as key quantum characteristics such as wave function parity and energy level distribution. Notably, the asymptotic behavior of the QBS levels aligns surprisingly well with ideal quantum systems, i.e., the one-dimensional finite and infinite square potential wells. We observed a stepwise increase in the number of QBS with increasing inter-barrier spacing, accompanied by reduced energy gaps between adjacent levels. Additionally, we identified critical thresholds essential for QBS existence and examined how variations in barrier width and height affect the number and position of these levels.
Using rectangular DBs as prototype systems, we explored the influence of boundary conditions and how the electronic configuration and optical characteristics of the well region can be finely tuned by adjusting inter-barrier spacing (the quasi-well width). The investigated DB systems exhibited atom-like optical absorption spectra, which are continuously adjustable through manipulation of inter-barrier spacing. Our findings highlight the intricate relationship between the geometric parameters of the barriers and their resultant electronic and optical properties, offering valuable insights into the design principles for nanostructures with tunable functionalities.
Moreover, by elucidating the free-atom-like characteristics of their electronic structures and optical responses, we demonstrated the potential of DB nanostructures as ultrahigh-sensitivity detectors for electromagnetic radiation. The ability to finely tune energy levels and optical responses within these nanostructures opens avenues for applications in ultrahigh-precision measurements, where accurate detection and analysis of light absorption at the nanoscale are critical.

\section*{Acknowledgments}

We acknowledge the financial support from National Natural Science Foundation of China (Grants No. 12074382, 11474285), and the Scientific Research Foundation for High-level Talents of Anhui University of Science and Technology under Grant YJ20240002. We are grateful to the Hefei Advanced Computing Center for support of supercomputing facilities. We also thank Professors Yugui Yao and Wenguang Zhu for their reading and helpful comments on the manuscript.

\appendix

\section{Derivation of dielectric function}\label{app:epsilon}

In this appendix, we derive the mathematical expression of the imaginary part of dielectric function, for the optical absorption of QBS. For the sake of simplicity, we consider the incident light wave as a monochromatic plane wave. The transition rate from the  $k$th energy level to the  $n$th energy level, which is the probability of transition per unit time, can be represented as \cite{zeng2007Quantum}
\begin{equation}
	\left|W_{nk}\right|=\frac{\pi}{6\hbar^{2}}\left|D_{nk}\right|^{2}E_{0}^{2}\delta(\omega_{n}-\omega_{k}-\omega),
\end{equation}
where $E_0$ is the vibrational amplitude of the corresponding electric field $E = E_0\cos(\omega t)$, $E_k = \hbar\omega_k$, $E_n = \hbar\omega_n$, are the $k$th energy level and the $n$th energy level, respectively. The transition matrix $|D_{nk}|^{2} \equiv |\langle k|-er|n\rangle|^{2}=e^{2}|\langle k|r|n\rangle|^{2}$ can be derived from the perturbation Hamiltonian $H^{'} = -e\phi = -erE_0\cos(\omega t) = D E_0\cos(\omega t) = W \cos(\omega t) $, with $D = -er$, involving the electron charge $e$ and the operator of real space coordinates $r$. It is evident that the parity (evenness or oddness) of the wave functions corresponding to the $k$th and $n$th energy levels determines whether the integral evaluates to zero or not. The transition rate can also be expressed as
\begin{widetext}
\begin{eqnarray}\label{eq:wnk}
	W_{nk}&=&\frac{\pi E_{0}^{2}}{6\hbar^{2}}|D_{nk}|^{2}\delta(\omega_{n}-\omega_{k}-\omega)
	 = \frac{\pi e^{2}E_{0}^{2}}{6\hbar^{2}}|\langle k|r|n\rangle|^{2}J_{nk}(\omega)
	=\frac{\pi e^{2}E_{0}^{2}}{6\hbar^{2}}A_{nk}(\omega).
\end{eqnarray}
\end{widetext}
Here, the transition rate $J_{nk}(\omega)=\delta(\omega_{n}-\omega_{k}-\omega)$ represents the distribution of energy levels corresponding to the transition energy difference $\hbar \omega$, which is associated with the joint density of states (JDOS) of the valence and conduction bands in semiconductor system.
The function $A_{nk}(\omega)=|\langle k|r|n\rangle|^{2}\delta(\omega_{n}-\omega_{k}-\omega)=|\langle k|r|n\rangle|^{2}J_{nk}(\omega)$ is determined by the energy level distribution and the transition matrix, reflecting the constraints of the selection rules on the light absorption spectrum of the electron bound states.

Next, we calculate the imaginary part of the dielectric function 
$\epsilon_{2}(\omega)$ from the perspective of medium absorption \cite{huang1998Solid}. For degenerate energy levels in the ground state, the absorption power per unit time is given by
\begin{equation}\label{eq:pnk}
	P_{nk} = 2\hbar\omega\times W_{nk}
\end{equation}
From the perspective of medium absorption, the absorption power per unit volume is
\begin{equation}
	\rho_{nk} =\sigma \bar{E^{2}} = 2\omega\epsilon_{2}(\omega)\epsilon_{0}E_0^2.
\end{equation}
In the potential well region, assuming the cross-sectional area of the double potential barrier is $S_0$, the corresponding volume $\Omega_0 = w S_0$. Then the absorption power is
\begin{equation}\label{eq:pnk_1}
	P_{nk} = \Omega_0\rho_{nk} = 2\Omega_0 \omega\epsilon_{2}(\omega)\epsilon_{0}E_0^2.
\end{equation} 
From Eqs.(\ref{eq:wnk}), (\ref{eq:pnk}) and (\ref{eq:pnk_1}), we have
\begin{equation}\label{eq:e2_1}
	2\hbar\omega\times \frac{\pi E_0^{2}}{6\hbar^2}|D_{nk}|^{2}\delta(\omega_{n}-\omega_{k}-\omega) = 2\Omega_0 \omega\epsilon_{2}(\omega)\epsilon_{0}E_0^2.
\end{equation}
Simplifying to get
\begin{equation}
	\epsilon_{2}(\omega)=\frac{\pi e^{2}}{6\hbar\epsilon_{0}\Omega_{0}}|D_{nk}|^{2}\delta(\omega_{n}-\omega_{k}-\omega).
\end{equation}
That is
\begin{eqnarray}
	\epsilon_{2}(\omega)&=&\frac{\pi e^{2}}{6\hbar\epsilon_{0}\Omega_{0}}|\langle k|r|n\rangle|^{2}\delta(\omega_{n}-\omega_{k}-\omega)
	\nonumber \\ &=&\frac{\pi e^{2}}{6\hbar\epsilon_{0}\Omega_{0}}|\langle k|r|n\rangle|^{2}J_{nk}(\omega).
\end{eqnarray}
This is a dimensionless quantity, directly corresponding to the experimentally observable light absorption properties of the DB nanostructure.

\section{Tables of Data}\label{app:table}
In this appendix, the data for initial state label $k=2$ and $k=3$ in Figs. \ref{fig:opab_11} and \ref{fig:opab_12} are listed in Tables \ref{tab:opab_21}-\ref{tab:opab_32}.

\begin{table*}[htbp]

	\caption{The optical absorption properties of DB system of electrons at different well widths $w = 10, \,20, \,30$ \AA, similar to Table \ref{tab:opab_11}. The initial state label $k=2$.}
	\label{tab:opab_21}
	\centering
	\resizebox{1.0\linewidth}{!}{

		\begin{tabular}{|c|c|c|c|c|c|c|c|c|c|c|c|c|}

			\hline
			$w$  & \multicolumn{4}{c|}{10} & \multicolumn{4}{c|}{20} & \multicolumn{4}{c|}{30}                                                                    \\
			\hline
			$n$ & $\hbar \omega$     & $\Gamma_{n}$      & $J_{nk}$        & $\epsilon_2$ & $\hbar \omega$ & $\Gamma_{n}$ & $J_{nk}$  & $\epsilon_2$ & $\hbar \omega$ & $\Gamma_{n}$ & $J_{nk}$  & $\epsilon_2$ \\
			\hline
			3  & 1.1342         & 4.4571$\times 10^{-5}$       & 8.9507$\times 10^{3}$       & 5.5547$\times 10^{3}$  & 0.3773     & 3.4509$\times 10^{-8}$  & 1.1560$\times 10^{7}$ & 1.0461$\times 10^{7}$  & 0.1812     & 3.3568$\times 10^{-9}$  & 1.1884$\times 10^{8}$ & 1.4869$\times 10^{8}$  \\
			\hline
			4  &             &             &             &       & 0.8979     & 4.3497$\times 10^{-7}$  & 9.1718$\times 10^{5}$ & ---     & 0.4340     & 1.2890$\times 10^{-8}$  & 3.0949$\times 10^{7}$ & ---     \\
			\hline
			5  &             &             &             &       & 1.5489     & 7.8591$\times 10^{-6}$  & 5.0762$\times 10^{4}$ & 4.1951$\times 10^{2}$  & 0.7570     & 6.9619$\times 10^{-8}$  & 5.7304$\times 10^{6}$ & 6.4477$\times 10^{4}$  \\
			\hline
			6  &             &             &             &       & 2.2942     & 2.9148$\times 10^{-4}$  & 1.3687$\times 10^{3}$ & ---     & 1.1481     & 4.4837$\times 10^{-7}$  & 8.8977$\times 10^{5}$ & ---     \\
			\hline
			7  &             &             &             &       &        &       &      &       & 1.6038     & 3.3706$\times 10^{-6}$  & 1.1836$\times 10^{5}$ & 1.2505$\times 10^{2}$  \\
			\hline
			8  &             &             &             &       &        &       &      &       & 2.1161     & 4.6616$\times 10^{-5}$  & 8.5580$\times 10^{3}$ & ---     \\
			\hline
			9  &             &             &             &       &        &       &      &       & 2.6613     & 9.3803$\times 10^{-4}$  & 4.2530$\times 10^{2}$ & 8.3380$\times 10^{-2}$  \\
			\hline
		\end{tabular}
	}
\end{table*}

\begin{table*}[htbp]
	\caption{The optical absorption properties of DB system of electrons at different well widths $w = 40,\, 50,\, 60$ \AA, similar to Table \ref{tab:opab_11}. The initial state label $k=2$.}
	\label{tab:opab_22}
	\centering
	\resizebox{1.0\linewidth}{!}{

		\begin{tabular}{|c|c|c|c|c|c|c|c|c|c|c|c|c|}

			\hline
			$w$  & \multicolumn{4}{c|}{40} & \multicolumn{4}{c|}{50} & \multicolumn{4}{c|}{60}                                                                    \\
			\hline
			$n$ & $\hbar \omega$     & $\Gamma_{n}$      & $J_{nk}$        & $\epsilon_2$ & $\hbar \omega$ & $\Gamma_{n}$ & $J_{nk}$  & $\epsilon_2$ & $\hbar \omega$ & $\Gamma_{n}$ & $J_{nk}$  & $\epsilon_2$ \\
			\hline
			3  & 0.1059         & 8.2558$\times 10^{-10}$       & 4.8323$\times 10^{8}$       & 7.7562$\times 10^{8}$  & 0.0693     & 2.8696$\times 10^{-10}$  & 1.3902$\times 10^{9}$ & 2.7268$\times 10^{9}$  & 0.0488     & 1.0581$\times 10^{-10}$  & 3.7703$\times 10^{9}$ & 8.7420$\times 10^{9}$  \\
			\hline
			4  & 0.2538         & 2.3578$\times 10^{-9}$       & 1.6920$\times 10^{8}$       & ---     & 0.1662     & 7.8456$\times 10^{-10}$  & 5.0849$\times 10^{8}$ & ---     & 0.1172     & 2.7503$\times 10^{-10}$  & 1.4505$\times 10^{9}$ & ---     \\
			\hline
			5  & 0.4436         & 8.4105$\times 10^{-9}$       & 4.7434$\times 10^{7}$       & 6.8183$\times 10^{5}$  & 0.2906     & 1.7978$\times 10^{-9}$  & 2.2191$\times 10^{8}$ & 3.8918$\times 10^{6}$  & 0.2049     & 7.2055$\times 10^{-10}$  & 5.5366$\times 10^{8}$ & 1.1471$\times 10^{7}$  \\
			\hline
			6  & 0.6748         & 3.3829$\times 10^{-8}$       & 1.1793$\times 10^{7}$       & ---     & 0.4425     & 4.7758$\times 10^{-9}$  & 8.3533$\times 10^{7}$ & ---     & 0.3121     & 2.0686$\times 10^{-9}$  & 1.9286$\times 10^{8}$ & ---     \\
			\hline
			7  & 0.9468         & 1.2283$\times 10^{-7}$       & 3.2479$\times 10^{6}$       & 4.3608$\times 10^{3}$  & 0.6216     & 1.5404$\times 10^{-8}$  & 2.5898$\times 10^{7}$ & 4.2328$\times 10^{4}$  & 0.4387     & 3.8962$\times 10^{-9}$  & 1.0239$\times 10^{8}$ & 1.9749$\times 10^{5}$  \\
			\hline
			8  & 1.2587         & 3.7955$\times 10^{-7}$       & 1.0511$\times 10^{6}$       & ---     & 0.8277     & 4.6237$\times 10^{-8}$  & 8.6283$\times 10^{6}$ & ---     & 0.5845     & 9.4192$\times 10^{-9}$  & 4.2354$\times 10^{7}$ & ---     \\
			\hline
			9  & 1.6090         & 2.0394$\times 10^{-6}$       & 1.9562$\times 10^{5}$       & 5.0937$\times 10^{1}$  & 1.0605     & 1.3269$\times 10^{-7}$  & 3.0067$\times 10^{6}$ & 9.5077$\times 10^{2}$  & 0.7495     & 2.0561$\times 10^{-8}$  & 1.9403$\times 10^{7}$ & 7.2308$\times 10^{3}$  \\
			\hline
			10 & 1.9952         & 1.0996$\times 10^{-5}$       & 3.6279$\times 10^{4}$       & ---     & 1.3195     & 3.4790$\times 10^{-7}$  & 1.1467$\times 10^{6}$ & ---     & 0.9335     & 5.0196$\times 10^{-8}$  & 7.9476$\times 10^{6}$ & ---     \\
			\hline
			11 & 2.4116         & 1.0379$\times 10^{-4}$       & 3.8439$\times 10^{3}$       & 2.7921$\times 10^{-1}$  & 1.6041     & 1.2754$\times 10^{-6}$  & 3.1280$\times 10^{5}$ & 2.7821$\times 10^{1}$  & 1.1363     & 1.1066$\times 10^{-7}$  & 3.6050$\times 10^{6}$ & 3.7740$\times 10^{2}$  \\
			\hline
			12 & 2.8392         & 1.4433$\times 10^{-3}$       & 2.7642$\times 10^{2}$       & ---     & 1.9131     & 5.4175$\times 10^{-6}$  & 7.3639$\times 10^{4}$ & ---     & 1.3577     & 2.9400$\times 10^{-7}$  & 1.3570$\times 10^{6}$ & ---     \\
			\hline
			13 &             &             &             &       & 2.2445     & 2.6259$\times 10^{-5}$  & 1.5192$\times 10^{4}$ & 4.7707$\times 10^{-1}$  & 1.5973     & 9.1721$\times 10^{-7}$  & 4.3495$\times 10^{5}$ & 1.6119$\times 10^{1}$  \\
			\hline
			14 &             &             &             &       & 2.5936     & 2.3329$\times 10^{-4}$  & 1.7101$\times 10^{3}$ & ---     & 1.8545     & 2.9447$\times 10^{-6}$  & 1.3548$\times 10^{5}$ & ---     \\
			\hline
			15 &             &             &             &       & 2.9443     & 2.7123$\times 10^{-3}$  & 1.4709$\times 10^{2}$ & 1.7495$\times 10^{-3}$  & 2.1285     & 1.3265$\times 10^{-5}$  & 3.0076$\times 10^{4}$ & 4.6143$\times 10^{-1}$  \\
			\hline
			16 &             &             &             &       &        &       &      &       & 2.4174     & 5.1916$\times 10^{-5}$  & 7.6844$\times 10^{3}$ & ---     \\
			\hline
			17 &             &             &             &       &        &       &      &       & 2.7173     & 3.7245$\times 10^{-4}$  & 1.0711$\times 10^{3}$ & 7.4866$\times 10^{-3}$  \\
			\hline
		\end{tabular}
	}
\end{table*}

\begin{table*}[htbp]
	\caption{The optical absorption properties of DB  system of electrons at different well widths $w = 10, \,20, \,30$ \AA, similar to Table \ref{tab:opab_11}. The initial state label $k=3$.}\label{tab:opab_31}
	\centering
	\resizebox{1.0\linewidth}{!}{

		\begin{tabular}{|c|c|c|c|c|c|c|c|c|c|c|c|c|}

			\hline
			$w$  & \multicolumn{4}{c|}{10} & \multicolumn{4}{c|}{20} & \multicolumn{4}{c|}{30}                                                                    \\
			\hline
			$n$ & $\hbar \omega$     & $\Gamma_{n}$      & $J_{nk}$        & $\epsilon_2$ & $\hbar \omega$ & $\Gamma_{n}$ & $J_{nk}$  & $\epsilon_2$ & $\hbar \omega$ & $\Gamma_{n}$ & $J_{nk}$  & $\epsilon_2$ \\
			\hline
			4  &             &             &             &       & 0.5205     & 4.3497$\times 10^{-7}$  & 9.1718$\times 10^{5}$ & 8.8337$\times 10^{5}$  & 0.2527     & 1.2890$\times 10^{-8}$  & 3.0949$\times 10^{7}$ & 4.0557$\times 10^{7}$  \\
			\hline
			5  &             &             &             &       & 1.1716     & 7.8591$\times 10^{-6}$  & 5.0762$\times 10^{4}$ & ---     & 0.5757     & 6.9619$\times 10^{-8}$  & 5.7304$\times 10^{6}$ & ---     \\
			\hline
			6  &             &             &             &       & 1.9169     & 2.9148$\times 10^{-4}$  & 1.3687$\times 10^{3}$ & 1.3609$\times 10^{1}$  & 0.9669     & 4.4837$\times 10^{-7}$  & 8.8977$\times 10^{5}$ & 1.1974$\times 10^{4}$  \\
			\hline
			7  &             &             &             &       &        &       &      &       & 1.4225     & 3.3706$\times 10^{-6}$  & 1.1836$\times 10^{5}$ & ---     \\
			\hline
			8  &             &             &             &       &        &       &      &       & 1.9349     & 4.6616$\times 10^{-5}$  & 8.5580$\times 10^{3}$ & 1.1992$\times 10^{1}$  \\
			\hline
			9  &             &             &             &       &        &       &      &       & 2.4801     & 9.3803$\times 10^{-4}$  & 4.2530$\times 10^{2}$ & ---     \\
			\hline
		\end{tabular}
	}
\end{table*}

\begin{table*}[htbp]
	\caption{The optical absorption properties of DB system of electrons at different well widths $w = 40, \,50, \,60$ \AA, similar to Table \ref{tab:opab_11}. The initial state label $k=3$.}\label{tab:opab_32}
	\centering
	\resizebox{1.0\linewidth}{!}{

		\begin{tabular}{|c|c|c|c|c|c|c|c|c|c|c|c|c|}

			\hline
			w  & \multicolumn{4}{c|}{40} & \multicolumn{4}{c|}{50} & \multicolumn{4}{c|}{60}                                                                    \\
			\hline
			$n$ & $\hbar \omega$     & $\Gamma_{n}$      & $J_{nk}$        & $\epsilon_2$ & $\hbar \omega$ & $\Gamma_{n}$ & $J_{nk}$  & $\epsilon_2$ & $\hbar \omega$ & $\Gamma_{n}$ & $J_{nk}$  & $\epsilon_2$ \\
			\hline
			4  & 0.1479         & 2.3578$\times 10^{-9}$       & 1.6920$\times 10^{8}$       & 2.8348$\times 10^{8}$  & 0.0969     & 7.8456$\times 10^{-10}$  & 5.0849$\times 10^{8}$ & 1.0398$\times 10^{9}$  & 0.0683     & 2.7503$\times 10^{-10}$  & 1.4505$\times 10^{9}$ & 3.5046$\times 10^{9}$  \\
			\hline
			5  & 0.3377         & 8.4105$\times 10^{-9}$       & 4.7434$\times 10^{7}$       & ---     & 0.2213     & 1.7978$\times 10^{-9}$  & 2.2191$\times 10^{8}$ & ---     & 0.1561     & 7.2055$\times 10^{-10}$  & 5.5366$\times 10^{8}$ & ---     \\
			\hline
			6  & 0.5689         & 3.3829$\times 10^{-8}$       & 1.1793$\times 10^{7}$       & 2.0170$\times 10^{5}$  & 0.3732     & 4.7758$\times 10^{-9}$  & 8.3533$\times 10^{7}$ & 1.7401$\times 10^{6}$  & 0.2633     & 2.0686$\times 10^{-9}$  & 1.9286$\times 10^{8}$ & 4.7424$\times 10^{6}$  \\
			\hline
			7  & 0.8409         & 1.2283$\times 10^{-7}$       & 3.2479$\times 10^{6}$       & ---     & 0.5523     & 1.5404$\times 10^{-8}$  & 2.5898$\times 10^{7}$ & ---     & 0.3899     & 3.8962$\times 10^{-9}$  & 1.0239$\times 10^{8}$ & ---     \\
			\hline
			8  & 1.1528         & 3.7955$\times 10^{-7}$       & 1.0511$\times 10^{6}$       & 1.8671$\times 10^{3}$  & 0.7584     & 4.6237$\times 10^{-8}$  & 8.6283$\times 10^{6}$ & 1.8614$\times 10^{4}$  & 0.5357     & 9.4192$\times 10^{-9}$  & 4.2354$\times 10^{7}$ & 1.0772$\times 10^{5}$  \\
			\hline
			9  & 1.5032         & 2.0394$\times 10^{-6}$       & 1.9562$\times 10^{5}$       & ---     & 0.9912     & 1.3269$\times 10^{-7}$  & 3.0067$\times 10^{6}$ & ---     & 0.7007     & 2.0561$\times 10^{-8}$  & 1.9403$\times 10^{7}$ & ---     \\
			\hline
			10 & 1.8894         & 1.0996$\times 10^{-5}$       & 3.6279$\times 10^{4}$       & 1.3503$\times 10^{1}$  & 1.2503     & 3.4790$\times 10^{-7}$  & 1.1467$\times 10^{6}$ & 5.1777$\times 10^{2}$  & 0.8846     & 5.0196$\times 10^{-8}$  & 7.9476$\times 10^{6}$ & 4.2242$\times 10^{3}$  \\
			\hline
			11 & 2.3057         & 1.0379$\times 10^{-4}$       & 3.8439$\times 10^{3}$       & ---     & 1.5348     & 1.2754$\times 10^{-6}$  & 3.1280$\times 10^{5}$ & ---     & 1.0875     & 1.1066$\times 10^{-7}$  & 3.6050$\times 10^{6}$ & ---     \\
			\hline
			12 & 2.7333         & 1.4433$\times 10^{-3}$       & 2.7642$\times 10^{2}$       & 2.8712$\times 10^{-2}$  & 1.8439     & 5.4175$\times 10^{-6}$  & 7.3639$\times 10^{4}$ & 9.9171$\times 10^{0}$  & 1.3088     & 2.9400$\times 10^{-7}$  & 1.3570$\times 10^{6}$ & 2.1499$\times 10^{2}$  \\
			\hline
			13 &             &             &             &       & 2.1752     & 2.6259$\times 10^{-5}$  & 1.5192$\times 10^{4}$ & ---     & 1.5484     & 9.1721$\times 10^{-7}$  & 4.3495$\times 10^{5}$ & ---     \\
			\hline
			14 &             &             &             &       & 2.5243     & 2.3329$\times 10^{-4}$  & 1.7101$\times 10^{3}$ & 8.4143$\times 10^{-2}$  & 1.8057     & 2.9447$\times 10^{-6}$  & 1.3548$\times 10^{5}$ & 7.9539$\times 10^{0}$  \\
			\hline
			15 &             &             &             &       & 2.8750     & 2.7123$\times 10^{-3}$  & 1.4709$\times 10^{2}$ & ---     & 2.0796     & 1.3265$\times 10^{-5}$  & 3.0076$\times 10^{4}$ & ---     \\
			\hline
			16 &             &             &             &       &        &       &      &       & 2.3686     & 5.1916$\times 10^{-5}$  & 7.6844$\times 10^{3}$ & 1.9295$\times 10^{-1}$  \\
			\hline
			17 &             &             &             &       &        &       &      &       & 2.6685     & 3.7245$\times 10^{-4}$  & 1.0711$\times 10^{3}$ & ---     \\
			\hline
		\end{tabular}
	}
\end{table*}

\newpage
\bibliography{ref.bib}

\end{document}